\documentclass[%
 reprint,
 superscriptaddress,
nofootinbib,
 amsmath,amssymb,
 aps,
]{revtex4-2}

\usepackage{graphicx}% Include figure files
\usepackage{dcolumn}% Align table columns on decimal point
\usepackage{xcolor} % Required for text coloring
\usepackage{bm}% bold math
\begin{document}

\preprint{APS/123-QED}

\title{A Linear Instability and Damping in the Acoustic Dispersion Relation of Fluids Subject to Inverse Compton Drag}% Force line breaks with \\

\author{Yiting Wang}
\homepage{ORCID: 0000-0002-3143-4552}
\email{wang2632@wisc.edu}
\affiliation{Department of Physics, University of Wisconsin-Madison}


\author{Sebastian Heinz}
\homepage{ORCID: 0000-0002-8433-8652}
\affiliation{Department of Physics, University of Wisconsin-Madison}
\affiliation{Department of Astronomy, University of Wisconsin-Madison}

\author{Vladimir Zhdankin}
\homepage{ORCID: 0000-0003-3816-7896}
\affiliation{Department of Physics, University of Wisconsin-Madison}

\date{Submitted: \today }% It is always \today, today,

\begin{abstract}
Radiation drag from inverse-Compton scattering of an external radiation field changes the acoustic dispersion relation of a relativistic fluid. We derive the linear dispersion relation and show that, depending on the electron distribution and fluid sound speed, radiation drag either damps both modes or damps one while driving the other unstable. The unstable growth is robust at short wavelengths but slow and overdamped at long wavelengths, where the linear analysis is limited by the background acceleration. The resulting growth and damping rates are of order the bulk radiative acceleration rate, so whenever radiation drag is important for the background flow, its effects on fluid perturbations are also important. A purely damped region exists for a wide range of parameters.  

Radiation drag also makes acoustic waves dispersive: The phase speed deviates from the ordinary sound speed at longer wavelengths, and approaches the sound speed at short wavelengths. We validate the derived rates using special-relativistic hydrodynamic simulations. We apply the derived relation to astrophysical jets such as gamma-ray bursts, blazars in the broad-line region of the host active galactic nucleus, and quasar jets traveling through the CMB and galactic radiation field. We find that the derived instability, damping and dispersion may be relevant across a range of relativistic outflow conditions.
\end{abstract}

\maketitle

%\tableofcontents

\section{\label{sec:intro}Introduction}
Relativistic flows are common in high-energy astrophysical systems, including Active Galactic Nuclei (AGN) jets, quasars, gamma-ray bursts, accretion disks, and pulsar winds. In most of the regions in these systems, the plasma propagates through an ambient radiation field supplied, for example, by an accretion disk, broad-line region (BLR), host galaxy, or the cosmic microwave background (CMB). The radiation field can interact with the relativistic particles via inverse-Compton scattering. In this paper, we focus on the  bulk momentum transfer between the radiation field and the plasma, commonly known as radiation drag or inverse Compton drag.
The radiation force due to scattering of the ambient radiation in the Thomson regime is given by \citet{phinney_acceleration_1982, sikora_radiation_1996}
\begin{equation}
    \frac{dp^{\mu}}{d\tau}
= -\sigma_T \left( u_\alpha R^{\alpha \mu} 
+ \left[ u_\alpha R^{\alpha \beta} u_\beta \right] u^{\mu}
\right),
\end{equation}
where $p^\mu$ is the particle four-momentum, $\tau$ is the proper time, $\sigma_T$ is the Thomson cross section, $u^\mu$ is the particle four-velocity, and $R^{\mu\nu}$ is the radiation stress-energy tensor. Depending on the distribution of the radiation field and the velocity of the plasma, the same radiation force can accelerate or decelerate the flow.

The first inverse-Compton dynamics was studied as Compton-rocket effect \citep{odell_radiation_1981,cheng_compton_1981,phinney_acceleration_1982} where the radiation field was shown to accelerate the relativistic hot plasma. 
\citet{odell_radiation_1981} showed that the Thomson-scattering radiation force on a relativistically hot plasma can be enhanced relative to that on a cold plasma by the anisotropic loss of internal energy due to an anisotropic radiation field.
\citet{cheng_compton_1981} emphasized that Compton losses can themselves drive bulk acceleration when the radiation field is sufficiently anisotropic and the relativistic enthalpy is carried by the scattering
particles.
\citet{phinney_acceleration_1982} pointed out that the same inverse-Compton interaction that provides the radiation force also produces strong radiative cooling. Thus, the Compton rocket is generally an inefficient way to convert internal random energy into bulk kinetic energy. The Compton-rocket force also generally drives the low-inertia plasma toward an equilibrium or terminal Lorentz factor \citep{renaud_terminal_1998} rather than providing unlimited acceleration. 

Instead, the same momentum-exchange force appears more as inverse-Compton radiation drag, extracting energy from the particles and decelerating the flow. 
\citet{begelman_inverse_1987} considered inverse-Compton scattering of ambient photons by a cold relativistic jet and showed how the transformation of the external radiation field into the comoving frame leads to anisotropic scattering, beamed emission, and momentum exchange with the flow.
\citet{melia_radiative_1989} studied radiative deceleration of relativistic jets by thermal radiation from an accretion disk. 
\citet{li_effects_1992} examined radiation drag in radial relativistic hydromagnetic winds.
\citet{sikora_radiation_1996} analyzed radiation drag in relativistic AGN jets containing both cold and relativistic particle components.
\citet{levinson_constraints_1996} showed strong radiative drag can constrain the pair content of relativistic jets.
\citet{luo_radiation_1999} extended the discussion of radiation forces in AGN jets to inverse-Compton scattering in both the Thomson and Klein-Nishina regimes.
Radiative deceleration has been studied in many high-energy blazars.
\citet{levinson_origin_2007} showed that an external radiation field can decelerate a relativistic blob. 
\citet{ghisellini_compton_2010} discussed the deceleration and recoil associated with anisotropic external-Compton emission in powerful blazars, emphasizing that strong radiative recoil is most important when the inertia of the flow is small. 

More recent work has considered Compton drag in dissipative relativistic flows, including blazar shells and Poynting-dominated outflows \citep{golan_blazar_2015,levinson_effect_2016}. In the context of gamma-ray bursts (GRBs) specifically, \citet{ceccobello_inverse-compton_2015} showed that inverse-Compton drag from the hot, optically thick cocoon surrounding a jet as it traverses the stellar envelope can decelerate a highly magnetized outflow to sub-relativistic speed, with an observable $1$-$100$ MeV flash as a signature of this process.
Radiation drag can also modify relativistic shock dynamics, as shown by \citet{leitus_dynamics_2017}. 
In optically thick flows, radiation can mediate the shock transition \citep{levinson_relativistic_2008}. Relativistic radiation-mediated shocks controlled by Compton scattering at high energies have been studied analytically and numerically \citep{levinson_relativistic_2008, levinson_physics_2020}. However, these radiation-mediated shocks are different from the optically thin external Compton drag problem considered in this paper.

Most previous studies have focused on the global effects of radiation drag force. In this paper, we instead study a local problem, where the inverse-Compton drag force changes the linear stability of a relativistic fluid when the force depends on the local thermodynamic state of a nonthermal electron population. Our calculation considers a small region of a relativistic plasma exposed to an external radiation field through an external inverse-Compton force term in the fluid equations. Because the force depends on the pressure and density through the moments of the electron distribution, small perturbations of the fluid variables also
perturb the force.

The purpose of this paper is to derive and analyze the local dispersion relation for a relativistic fluid subject to inverse-Compton drag.
In Section \ref{sec:linear_analysis}, we set up the problem in 1D and write the radiation drag force for an initially isotropic distribution of relativistic particles in the linearized hydrodynamic equations.
In Section \ref{sec:analytical_results}, we derive the analytical solution for the dispersion relations.
In section \ref{sec:numerical_results}, we use special relativistic hydrodynamic simulations to confirm the (in)stability in the linear regime.
In Section \ref{sec:discussion}, we discuss the astrophysical implications of this work and the limitations. Finally, in Section \ref{sec:conclusion}, we present a summary of our conclusion.

\section{\label{sec:linear_analysis}Physical Model and Linearized Equations}
Throughout this work we consider a plane-parallel radiation field propagating along $+\hat{x}$. This approximation is relevant in two general physical regimes: (a) a fluid subject to a plane-parallel radiation field from a "distant" source (in which case the fluid can be at rest and still experience the effects) and (b) a fluid moving at relativistic speed through some background radiation field, so that a field which is genuinely isotropic in the source frame appears effectively unidirectional in the comoving frame. The transformation between comoving energy density $U'$ and lab frame $U$ differs in the two regimes. We will discuss both boost transform examples in Section \ref{sec:discussion}. Our dispersion analysis is based in the initial rest frame of the fluid, so it will apply to both cases under the condition that the effect is evaluated locally and in a short time. We discuss more about this in Section \ref{sec:limitations}.

\subsection{\label{sec:radiation_force}Radiation-drag force}
First, we derive the radiation-drag force from the plane-parallel radiation field in the lab frame to the fluid rest frame directly from \citet{phinney_acceleration_1982}.
In the lab frame, its stress-energy tensor is
\begin{equation}
R^{\mu\nu}=U n^\mu n^\nu,
\qquad
n^\mu=(1,1,0,0),
\end{equation}
where $U$ is the lab-frame radiation energy density.

The fluid moves along the $\hat{x}$ direction with velocity in the lab frame 
\begin{equation}
\boldsymbol{\beta}(t)=(\beta(t),0,0),
\ \ \ 
u^\mu=\Gamma(1,\boldsymbol{\beta}),
\ \ \ 
\Gamma=(1-\beta^2)^{-1/2}.
\end{equation}
Thus, $\beta<0$ corresponds to motion along -$\hat{x}$.
Let the fluid have initial lab-frame velocity $\beta_0$, with
\begin{equation}
\Gamma_0=(1-\beta_0^2)^{-1/2}.
\end{equation}
In the fluid's initial rest frame, an electron has random velocity $\boldsymbol{w}$ and four-velocity
\begin{equation}
u_e^{\prime\mu}=\gamma_e(1,\boldsymbol{w}),
\qquad
\gamma_e=(1-\boldsymbol{w}^2)^{-1/2}.
\end{equation}
The radiation field in the fluid's initial rest frame has stress-energy tensor $R'^{\mu \nu}$
\begin{equation}
R^{\prime\mu\nu}
= \Lambda^{\mu\alpha} R_{\alpha\beta} \Lambda^{\beta\nu}
=U'k^\mu k^\nu,
\qquad
k^\mu=(1,1,0,0),
\end{equation}
where $\Lambda^{\mu\alpha}$ is the Lorentz transformation from the lab frame to the fluid's initial rest frame, with
\begin{equation}
U'=U\Gamma_0^2(1-\beta_0)^2,
\qquad
\Gamma_0=(1-\beta_0^2)^{-1/2},
\end{equation}
where $U'$ is the radiation energy density measured in the fluid initial rest frame.

In the fluid's frame, one electron feels the 4-force from this $R'^{\mu \nu}$: 
\begin{equation}
    \frac{dp'^{\mu}}{d\tau'}
= -\sigma_T \left\{
u'_\alpha R'^{\alpha \mu} 
+ \left[ u'_\alpha R'^{\alpha \beta} u'_\beta \right] u'^{\mu}
\right\}.
\end{equation}
The 4-force term in the fluid's rest frame is
\begin{align}
\frac{dp'^{0}}{d\tau'}
& = -\sigma_T U' \gamma_e (w_x - 1)\left[
1 + \gamma_e^2 (w_x - 1)
\right],\\
\frac{dp'^{1}}{d\tau'}
& = -\sigma_T U' \gamma_e (w_x - 1)\left[
1 + \gamma_e^2 w_x (w_x - 1)
\right],\\
\frac{dp'^{2}}{d\tau'}
& = -\sigma_T\,U'\,\gamma_e^3\,w_y\,(w_x - 1)^2,\\
\frac{dp'^{3}}{d\tau'}
& = -\sigma_T\,U'\,\gamma_e^3\,w_z\,(w_x - 1)^2.
\end{align}
In fluid's rest frame time: 
$\frac{dp'^{\mu}}{dt'}
= \frac{1}{\gamma_e}\frac{dp'^{\mu}}{d\tau'}$. Assuming the electron's velocity distribution is isotropic in the fluid's initial rest frame, \footnote{The anisotropic force would distort this distribution away from isotropy, but only on the radiative-loss timescale. Since our analysis is local in time, we may take the distribution isotropic at the initial instant. See \ref{sec:nocoolingcaveat}.}
\begin{align}
\langle w_i \rangle &= 0, \\
\langle w_i w_j \rangle &= \tfrac{1}{3}\langle w^2 \rangle \delta_{ij}, \\
\langle \gamma_e w_i w_j \rangle &= \tfrac{1}{3}\langle \gamma_e w^2 \rangle \delta_{ij},\\
\langle \gamma_e^2 w_i w_j \rangle &= \tfrac{1}{3}\langle \gamma_e^2 w^2 \rangle \delta_{ij}.
\end{align}
The force terms on the particle distribution in the rest frame of the fluid are given by
\begin{align}
f^0 = \left\langle n_e \frac{dp'^{0}}{dt'} \right\rangle
& = n_e \sigma_T U'
   \left[1 - \langle \gamma_e^2 \rangle 
   - \frac{1}{3}\langle \gamma_e^2 w^2 \rangle \right],
   \label{eq:energy_term_force}
\\
f^1 = \left\langle n_e \frac{dp'^{1}}{dt'} \right\rangle
&= n_e \sigma_T U'
   \left[1 + \frac{2}{3}\langle \gamma_e^2 w^2 \rangle \right].
\end{align}
The third and fourth terms vanish. All the averaged terms $\langle Q \rangle$ depend on the electron energy distribution.

\subsection{\label{sec:linearized_equations}Linearized relativistic hydrodynamic equations}
Throughout the linearized derivation we adopt $c=1$. 
We choose the fluid initial rest frame $S_0$ as the frame in which the hydrodynamic equations are linearized. All quantities are measured in $S_0$. In this frame, the unperturbed fluid is initially at rest,
\begin{equation}
\beta'_0(t=0)=0,
\qquad
\Gamma'_0(t=0)=1.
\end{equation}
However, because the background radiation force may accelerate the
fluid, we allow
\begin{equation}
\left.\frac{\partial \beta'_0}{\partial t}\right|_{t=0}\neq 0.
\end{equation}

In 1D special relativistic hydrodynamics,\footnote{We restrict to perturbations whose gradients lie along the direction of the bulk motion and the radiation field; oblique and transverse modes are outside the present treatment. See \ref{sec:onedimcaveat}.}the continuity equation and momentum equation are
\begin{align}
\frac{\partial}{\partial t}(\rho \Gamma')+\frac{\partial}{\partial x}(\rho \Gamma' \beta') & =0,\\
\frac{\partial}{\partial t}(\Gamma'^2 w \beta') +\frac{\partial}{\partial x}(\Gamma'^2 w \beta'^2+P) & =f^1,
\end{align}
where the relativistic enthalpy density is
\begin{equation}
w=\rho+\frac{\Gamma_{\rm adi}}{\Gamma_{\rm adi}-1}P.
\end{equation}
We adopt the pure-drag closure $f^0=0$ and no comoving-frame energy exchange, so the radiation transfers momentum to the flow without heating it.

After linearization
\begin{align}
\rho & =\rho_0+\delta \rho, \quad \beta'=\beta'_0+\delta \beta', \\ P & = P_0+\delta P, \quad f^1  =f^1_{(0)}+\delta f^1,
\label{eq:pertubation}
\end{align}
%under the assumptions of 
%\begin{equation}
%\beta'_0=0, \quad \Gamma'_0=1, \quad 
%\frac{\partial \rho_0}{\partial x}=0, \quad 
%\frac{\partial \beta'_0}{\partial x}=0, \quad 
%\frac{\partial \rho_0}{\partial t}=0
%\end{equation}
%\begin{equation}
%\frac{\partial \beta'_0}{\partial t}\neq 0
%\end{equation}
keeping \(f^1_{(0)}\) and \(\delta f^1\) explicit, the linearized relativistic equations in 1st order are
\begin{align}
\frac{\partial (\delta\rho)}{\partial t}
+\rho_0\frac{\partial (\delta\beta')}{\partial x}
+\rho_0\frac{\partial \beta'_0}{\partial t}\,\delta\beta' & =0, \\
w_0\frac{\partial (\delta\beta')}{\partial t}
+\left(
\delta\rho+\frac{\Gamma_{\rm ad}}{\Gamma_{\rm ad}-1}\delta P
\right)\frac{\partial \beta'_0}{\partial t}
+\frac{\partial (\delta P)}{\partial x}
 & =\delta f^1,
\end{align}
with 
\begin{equation}
w_0\frac{\partial \beta'_0}{\partial t}=f^1_{(0)},
\end{equation}
and
\begin{equation}
    w_0=\rho_0 + \frac{\Gamma_{ad}}{\Gamma_{ad}-1}P_0.
    \label{eq:background_acceleration}
\end{equation}
The background acceleration $\partial\beta'_0/\partial t=f^1_{(0)}/w_0$ is kept whereas $\partial w_0/\partial t$ vanishes under the pure-drag closure and is neglected. This holds provided radiative cooling is slow compared with the growth rate, See Section \ref{sec:nocoolingcaveat}.

We define the dimensionless enthalpy as
\begin{equation}
    \tilde{w}_{0}\equiv \left(1 + \frac{\Gamma_{\rm ad}}{\Gamma_{\rm ad}-1}\frac{P_0}{\rho_0}\right),
\end{equation}
then $w_0=\rho_0 \tilde{w}_{0}$. For use below, we also define the protonization parameter
\begin{equation}
    {\chi}\equiv \frac{\rho_{\rm e,0}}{\rho_0}=\frac{\rho_{\rm e}}{\rho_{\rm e}+\rho_{\rm p}}=\frac{1}{1 + \frac{\rho_{\rm p}}{\rho_{\rm e}}}
    \label{eq:chi}
\end{equation}
where $\rho_{\rm e}$ and $\rho_{\rm p}$ are the lepton and proton rest mass density, respectively, with $m_{\rm e}/m_{\rm p}\leq\chi\leq1$, then it gives $n_{\rm p}\leq n_{\rm e}$.

Finally, it is useful to define the dimensionless radiation drag efficiency parameter. From here on we write the electron velocity average $\langle\gamma_{\rm e}^2 w^2\rangle_0\equiv\langle\gamma^2\beta^2\rangle_0$, dropping the subscript $\rm e$ since all averages are over the electron distribution in the unperturbed background,
\begin{equation}
    \tilde{\chi} \equiv \left[{1 + \frac{2}{3}\langle\gamma^2\beta^2\rangle_0}\right]{\chi}
    \label{eq:tildechi}
\end{equation}
which measures the relative importance of the relativistic drag and the inertial density of protons in the problem.

\section{\label{sec:analytical_results}Analytic Results}

\subsection{\label{sec:dispersion_relation}Dispersion relation}
From
\begin{equation}
    f^{1} = n_{\rm e}\sigma_{\rm T}U'\left[1 + \frac{2}{3}\langle\gamma^2\beta^2\rangle\right],
\end{equation}
we first calculate the perturbation to the force term:
\begin{equation}
    \delta f^{1}=f_{0}\left[\frac{\delta U'}{U'_0} +\frac{\delta \rho}{\rho_{0}} + \frac{2}{3}\frac{\delta \rho}{\rho_{0}}\xi \right],
\end{equation}
where we define
\begin{equation}
    \xi\equiv \frac{2}{3}\frac{n_0\sigma_T U'_0\langle\gamma^2\beta^2\rangle_0}{f_0} = \frac{2\langle\gamma^2\beta^2\rangle_0}{3 + 2\langle\gamma^2\beta^2\rangle_0},
    \label{eq:xi_definition}
\end{equation}
which satisfies $0\leq\xi\leq1$. Physically, $\xi$ measures the fractional contribution of the electrons' random relativistic motion to the total radiation force. For sub-relativistic electrons, $\langle\gamma^2\beta^2\rangle_0\ll1$ and $\xi\ll1$, while for highly relativistic electrons $\xi\rightarrow1$. We also define the unperturbed force density as
\begin{equation}
    f_{0}\equiv n_{e,0}\sigma_{\rm T}U'_0\left[1 + \frac{2}{3}\langle\gamma^2\beta^2\rangle_0\right].
    %\equiv \tilde{f}_0\left[1 + \frac{2}{3}\langle\gamma^2\beta^2\rangle_0\right].
\end{equation}
%with $\tilde{f}_0\equiv n_{\rm e,0}\sigma_{\rm T}U'_0$.

With
\begin{equation}
    \delta U'=-2\delta \beta U'_0,
\end{equation}
we have the first order term of the force density:
\begin{equation}
    \delta f^{1}=f_{0}\left[-2\delta\beta + \left(1 + \frac{2\xi}{3}\right)\frac{\delta \rho}{\rho_{0}}\right].
    \label{eq:perturbed_force}
\end{equation}
This casts the perturbed fluid equations as
\begin{equation}
    \frac{\partial \delta \rho}{\partial t} + \rho_{0}\frac{\partial \delta \beta}{\partial x} + \rho_{0}\frac{\partial \beta_{0}}{\partial t}\delta \beta = 0,
    \label{eq:perturbed_continuity}
\end{equation}
\begin{equation}
    w_{0}\frac{\partial \delta \beta}{\partial t} + \left(\delta\rho + \frac{\Gamma_{ad}}{\Gamma_{ad}-1}\delta P\right)\frac{\partial \beta_0}{\partial t}+\frac{\partial \delta P}{\partial x}=\delta f^{1},
\label{eq:perturbedmomentum}
\end{equation}
\begin{equation}
    \delta P=\Gamma_{ad}\frac{P_0}{\rho_0}\delta\rho.
    \label{eq:perturbned_state}
\end{equation}
%\begin{equation}
%    w_0\frac{\partial \beta_0}{\partial t}=f^1_0
%\end{equation}
Inserting Eqs.~(\ref{eq:background_acceleration}), (\ref{eq:perturbed_force}), and (\ref{eq:perturbned_state}) into the momentum equation yields
\begin{multline}
    w_{0}\frac{\partial \delta \beta}{\partial t}
    + \left(\delta\rho + \Gamma_{\rm ad}\,\frac{\Gamma_{\rm ad}}{\Gamma_{\rm ad}-1}\frac{P_0}{\rho_0}\,\delta \rho\right)\frac{f_0}{w_0}
    + \Gamma_{\rm ad}\frac{P_0}{\rho_0}\frac{\partial \delta \rho}{\partial x} \\
    = f_0\left[-2\delta\beta + \left(1+\frac{2\xi}{3}\right)\frac{\delta \rho}{\rho_0}\right].
\end{multline}
We analyze the perturbations locally in time about the initial background state and adopt the Fourier decomposition
\begin{equation}
    \delta q(x,t) = \delta q_k \exp[i(kx-\omega t)],
\end{equation}
for each perturbed quantity $\delta q$. Under this decomposition,
\begin{equation}
    \frac{\partial}{\partial t} \rightarrow -i\omega,
    \qquad
    \frac{\partial}{\partial x} \rightarrow ik.
\end{equation}
%Although the radiation force accelerates the unperturbed flow, we assume that the background quantities vary slowly over the local time interval used to define the dispersion relation. We therefore treat
%\begin{equation}
%    \frac{f_0}{w_0} \equiv \omega_0
%\end{equation}
%as a constant background coefficient. It passes through the Fourier transform unchanged.

Applying this to the continuity equation gives
\begin{equation}
    -i\omega \delta \rho + \rho_0 ik\delta \beta + \rho_0\frac{f_0}{w_0}\delta\beta = 0,
\end{equation}
which we rewrite as
\begin{equation}
    \delta \rho=\frac{\left(ik + \frac{f_0}{w_0}\right)}{i\omega}\rho_0\delta\beta.
    \label{eq:deltarho_deltabeta_relation}
\end{equation}
Note, $\delta \rho$ is close to in phase with $\delta \beta$ when k is large, while it may be substantially out of phase when k is small.

Fourier transform of the momentum equation gives:
\begin{align}
    -w_0\,i\omega\,\delta\beta
    & + \left(1 + \Gamma_{\rm ad}\,\frac{\Gamma_{\rm ad}}{\Gamma_{\rm ad}-1}\frac{P_0}{\rho_0}\right)\frac{f_0}{w_0}\delta\rho
    + \Gamma_{\rm ad}\,ik\,\frac{P_0}{\rho_0}\delta\rho \notag\\
    &= f_0\left[\left(1+\frac{2\xi}{3}\right)\frac{\delta\rho}{\rho_0}-2\delta\beta\right].
    \label{eq:momentumlinear}
\end{align}
%We substitute $\delta\rho$ from the %continuity equation:
%\begin{align}
%    & \left(-w_0 i\omega + 2f_0\right)\delta\beta \\
%    & + \left[\left(1 + \frac{\Gamma_{\rm ad}^2}{\Gamma_{\rm ad}-1}\frac{P_0}{\rho_0}\right)\frac{f_0}{w_0} + \Gamma_{\rm ad}\,ik\,\frac{P_0}{\rho_0} - \left(1+\frac{2\xi}{3}\right)\frac{f_0}{\rho_0}\right]\delta\rho \\
%    & = 0
%\end{align}

We substitute $\delta\rho$ from the continuity equation
%\begin{align}
%    \big(- & w_0  i\omega  + 2f_0\big) \delta\beta \notag \\
%    + & \left[\left(1 + \frac{\Gamma_{\rm ad}^2}{\Gamma_{\rm ad}-1}\frac{P_0}{\rho_0}\right)\frac{f_0}{w_0} + \Gamma_{\rm ad}\,ik\,\frac{P_0}{\rho_0} - \left(1+\frac{2\xi}{3}\right)\frac{f_0}{\rho_0}\right] \notag \\
%    \times & \frac{\left(ik + \dfrac{f_0}{w_0}\right)\rho_0}{i\omega}\,\delta\beta = 0
%\end{align}
and solve for the dispersion relation% (start by multiplying with $i\omega$ and dividing by $\delta\beta$):
\begin{align}
     \omega^2 & + 2i\omega\frac{f_0}{w_0}\notag \\
    & + \bigg \{ \left[-\frac{2\xi}{3}\rho_0 + \left(1-\frac{2}{3}\frac{\xi}{\Gamma_{\rm ad}-1}\right)\Gamma_{\rm ad}P_0\right]\frac{f_0}{w_0}\notag \\
    & \ \ \ \ \ +\Gamma_{\rm ad}\,ik\,P_0 \bigg \} \left(\frac{ik}{w_0} + \frac{f_0}{w_0^2}\right)=0,
    \label{eq:dispersion_relation}
\end{align}
with roots
\begin{align}
     \omega_{\pm} = &\omega_0\Bigg[-i \label{eq:dispersion_roots} \\ & \ \ \ \pm \left.\sqrt{ -1 + \frac{2\xi}{3} + a_{s}^2\left(\frac{k^2}{\omega_0^2} - 1\right) +\frac{ik}{\omega_0}\left(\frac{2\xi}{3}-2a_s^2\right)}\right]\notag,
\end{align}
where we define the characteristic frequency of the problem as 
\begin{equation}
    \omega_0 \equiv \frac{f_0}{w_0}
\end{equation}
which is equal to the initial acceleration rate of the background flow. The relativistic adiabatic sound speed $a_{\rm s}$ is defined through
\begin{equation}
    a_{s}^2 \equiv \Gamma_{ad}\frac{P_0}{w_0} \leq \frac{1}{3}
\end{equation}
 
In the absence of radiation drag, $f_0\rightarrow0$, the linearized relativistic hydrodynamic equations recover the ordinary acoustic dispersion relation
\begin{equation}
    \omega^2=a_s^2k^2,
\end{equation}
or equivalently
\begin{equation}
    \omega_\pm=\pm a_s k.
\end{equation}

\subsection{\label{sec:stability_analysis}Stability analysis}
We can find the imaginary and real parts of $\omega$ as
\begin{align}
    \omega_i &= \omega_0 \mathrm{ sign}(b)\left[-1 \pm \sqrt{\frac{z-a}{2}}\right],\\
    \omega_r &= \omega_0 \sqrt{\frac{z+a}{2}},
\end{align}
where we define
\begin{align}
    a &\equiv -1 + \frac{2\xi}{3} + a_s^2\left(\frac{k^2}{\omega_0^2}-1\right),
    \label{eq:a}\\
    b &\equiv \frac{k}{\omega_0}\left(\frac{2\xi}{3} -2 a_s^2\right),
    \label{eq:b}\\
    z &\equiv \sqrt{a^2 + b^2}.
\end{align}
The $\mathrm{sign}(b)$ determines which of the left or right moving waves have the larger imaginary roots (see also Section \ref{sec:discussion}). 

Whether a mode grows or is damped depends on the sign of the imaginary roots. The boundaries between growing and damped modes are obtained by requiring marginal stability,
\begin{equation}
    \omega_i = \mathrm{Im}(\omega)=0.
\end{equation}
Combining this condition with Eqs.~(\ref{eq:a}) and (\ref{eq:b}) gives the critical value of $\xi$ as a function of wavenumber,
%\begin{eqnarray*}
%    & & \omega_i =  0 \\
%    & \Longleftrightarrow & -1 \pm \sqrt{\frac{z-a}{2}} = 0 \\
%    & \Longleftrightarrow & 2 = \sqrt{a^2+b^2} - a \\
%    & \Longleftrightarrow & b^2 - 4 - 4a = 0
%\end{eqnarray*}
%We can insert the definitions for $a$ and $b$ from above to get
%\begin{equation}
%    \frac{4}{9}\frac{k^2}{\omega_0^2}\xi^2 - \frac{8}{3}\frac{k^2}{\omega_0^2}\xi a_s^2 + 4 \frac{k^2}{\omega_0^2}a_s^4 - \frac{8}{3}\xi -4\frac{k^2}{\omega_0^2}a_s^2 + 4 a_s^2 = 0 
%\end{equation}
which yields
\begin{equation}
    \xi_0(k) =3\left[a_s^2 +\frac{\omega_0^2}{k^2}\left(1 \pm \sqrt{1 + \frac{a_s^2k^2}{\omega_0^2} + \frac{a_s^2 k^4}{\omega_0^4}}\right)\right].
\end{equation}

In the long-wavelength limit, 
\begin{equation}
    \xi_{\rm low} \equiv \lim\limits_{k\rightarrow 0}\xi_0=\frac{3}{2}a_s^2.
    \label{eq:xilow}
\end{equation}
In the short-wavelength limit, the two formal solutions approach
\begin{equation}
    \lim\limits_{k\to \infty}\xi_0=3\left[a_s^2 \pm \left(a_s + \frac{a_{s}^2}{2}\right)\right].
    \label{eq:large_k_limit}
\end{equation}
Since $a_s < 1$ and we require $0 \leq \xi \leq 1$, we must choose the positive root to find
\begin{equation}
    \xi_{\rm high}  \equiv  \lim\limits_{k\to \infty}\xi_0=3\left(\frac{a_s^2}{2} + a_s\right).
    \label{eq:xihigh}
\end{equation}

For $\xi < \xi_{\rm low}$, growing modes exist at long wavelengths, while for $\xi > \xi_{\rm high}$, growing modes exist at short wavelengths.  

We note that for the ultra-relativistic case with $a_{\rm s}^2 \rightarrow 1/3$ and $\xi \rightarrow 1$, no growing modes exist, since $\xi_{\rm high} \rightarrow 1 + \sqrt{3} > 1$ and $\xi > \xi_{\rm low} = \frac{1}{2}$.

The limiting behavior of $\omega_{i}$ can be found as
\begin{equation}
    \lim\limits_{k\to\infty}\omega_i = -\omega_0\left\{1 \pm\ a_s\left[1 - \xi/\left(3a_s^2\right)\right]\right\}
    \label{eq:omega_i_at_k_infty},
\end{equation}
and
\begin{equation}
    \lim\limits_{k\to 0}\omega_i=-\omega_0\left(1 \pm \sqrt{1+a_s^2-2\xi/3}\right),
    \label{eq:omega_i_at_k_0}
\end{equation}
consistent with the limits on $\xi$ derived above.

\begin{figure}
    \begin{center}  
    \includegraphics[width=\linewidth]{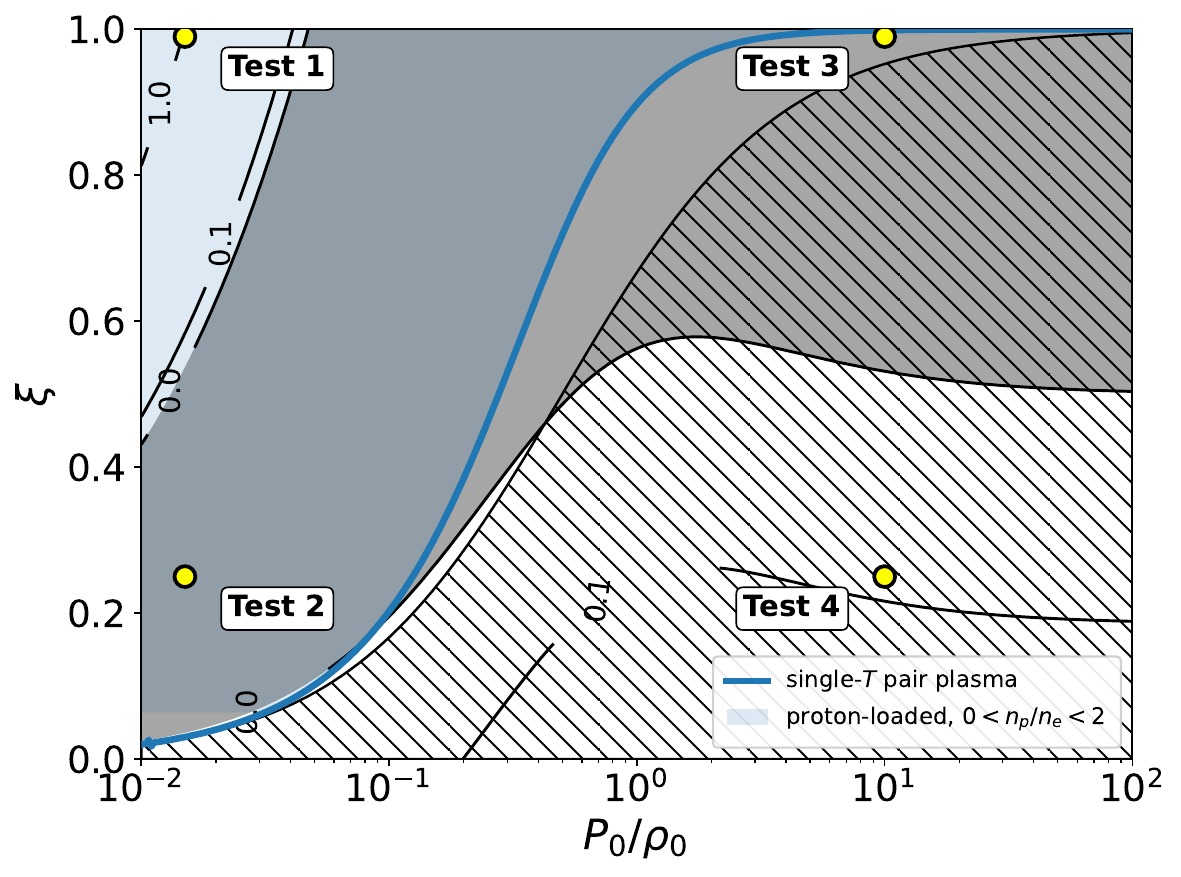}
    \end{center}
    \caption{Maximum growth rate as a function of $\xi$ and $P_0/\rho_0$. Here, we chose $\Gamma=((4/3)P_0 + (5/3)\rho_0)/(P_0+\rho_0)$ to approximate the change in adiabatic index for cold and hot plasmas. The gray shaded area shows the region of parameter space that is damped for both modes. The hatched area shows the lepton-forbidden region $\xi < \xi_{\rm min}$ (note that this region is not strictly prohibited if the pressure is dominated by a hadronic component).\footnote{Here we assume that the leptonic and hadronic components remain sufficiently coupled to share a common bulk velocity. Differential radiative acceleration between species can itself drive kinetic instabilities when this approximation fails \citep{vanthieghem_role_2022, levinson_anomalous_2023}. Resolving this regime requires a multi-fluid or kinetic treatment and lies outside the present model.} The solid blue curve traces the states of a single-temperature Maxwell-Jüttner pair plasma ($P_0/\rho_0=\Theta$, $\xi=\xi(\Theta)$, with $\Theta=kT/m_ec^2$), and the light-blue band shows how proton loading ($0<n_p/n_e<2$) extends these to lower $P_0/\rho_0$ at fixed $\xi$. Yellow circles mark the three parameter points used for the linear-wave validation runs of Table~\ref{tab:wave_cases}: Tests~1 and~4 sample the two unstable regions, while Tests~2 and~3 lie in the damped region. Test 4 is in the short-wavelength unstable region which will be discussed more in Section~\ref{sec:discussion}}
    \label{fig:max_rate}
\end{figure}

Finally, the electron distribution imposes an additional constraint on the accessible values of $\xi$. Since $\gamma\geq1$,
\begin{equation}
    \left\langle\gamma^2\beta^2\right\rangle
    \geq
    \left\langle\gamma\beta^2\right\rangle .
\end{equation}
For an isotropic electron population, the electron pressure is
\begin{equation}
    P_e = \frac{1}{3}\rho_e \left\langle\gamma\beta^2\right\rangle .
\end{equation}
Since the total rest-mass density satisfies $\rho\geq\rho_e$, we also have $P_e/\rho_e\geq P_e/\rho$.
Because $\xi$ is a monotonic function of
$\langle\gamma^2\beta^2\rangle$, these relations give the lower bound
\begin{equation}
    \xi \geq \xi_{\rm min} \equiv \frac{2P_{e,0}/\rho_0}{1 + 2P_{e,0}/\rho_0} .
\end{equation}
For a purely leptonic flow, $P_0=P_{e,0}$, so $\xi_{\rm min}$ becomes a direct function of $P_0/\rho_0$. 
The corresponding excluded region,
$\xi > \xi_{\rm min}$, is shown by the hatched area in Fig.~\ref{fig:max_rate}. This constraint removes the low-$\xi$ unstable branch for a purely leptonic plasma. The lower branch can become accessible if the fluid pressure is carried by a component that does not contribute directly to the inverse-Compton force, for example, in a lepton-hadron plasma, $P_0=P_{e,0}+P_{h,0}$, which in principle, is possible if lepton cooling has strongly suppressed the lepton pressure. 

Fig.~\ref{fig:max_rate} summarizes the stability properties in the $(P_0/\rho_0,\xi)$ plane. At each point, we maximize $\mathrm{Im}(\omega)$ over wavenumber and over the two roots of the dispersion relation. The gray region corresponds to parameters for which both modes are damped at all wavelengths. The white region at high $\xi$ contains the short-wavelength instability, while the white region below the gray boundary contains the long-wavelength, low-$\xi$ instability. The hatched region shows the excluded region $\xi<\xi_{\rm min}$ for a purely leptonic plasma. Tests~1 and~4 therefore sample the two unstable branches, whereas Tests~2 and~3 lie in the region that is damped for all wavelengths. Test~4 lies inside the purely leptonic exclusion region and is used as a numerical test of the low-$\xi$ branch rather than as a model of a purely leptonic thermal plasma.

The stability plot Fig.~\ref{fig:max_rate} is a general stability map where we vary $P_0/\rho_0$ and $\xi$ independently to allow for different plasma compositions and electron distributions.
For a single-temperature electron-positron plasma with a Maxwell-Jüttner distribution, the dimensionless temperature $\Theta=kT/m_ec^2$ sets both axes: $P_0/\rho_0=\Theta$ from the ideal-gas law, while the same temperature fixes $\langle\gamma^2\beta^2\rangle(\Theta)$ and hence $\xi$
\begin{equation}
\langle\gamma^2\beta^2\rangle(\Theta)= \frac{\int_1^\infty(\gamma^2-1)\,\gamma^2\beta\,e^{-\gamma/\Theta}\,d\gamma}{\int_1^\infty\gamma^2\beta\,e^{-\gamma/\Theta}\,d\gamma}.
\end{equation}

The accessible states then follow a one-parameter curve (solid blue in Fig.~\ref{fig:max_rate}), which is close to the stability boundary. A thermal pair plasma is unstable only in a narrow, weakly growing segment at low temperature ($\Theta\lesssim0.07$, rates below $\sim10^{-3}\omega_0$). If we include cold protons with $\mu=n_p/n_e$, we expand the parameter space to $P_0/\rho_0=\Theta_e/(1+\mu\,m_p/m_e)$ (light blue region).

\section{\label{sec:numerical_results}Numerical Results}

\subsection{\label{sec:numerical_setup}Numerical setup}
We use the {\tt Athena++} code  \citep{stone_athena_2008} to verify the linear stability analysis developed in \S~\ref{sec:dispersion_relation} against fully nonlinear simulations. The relativistic mode of Athena++ evolves the equations of special relativistic hydrodynamics in conservative form \citep{stone_athena_2020}
\begin{align}
\partial_t D + \partial_j(\Gamma\rho v^j) &= 0, \\
\partial_t E + \partial_j M^j &= G^0, \\
\partial_t M^i + \partial_j\!\left(\Gamma^2 w_{\rm gas} v^i v^j 
+ P\,\delta^{ij}\right) &= G^i,
\end{align}
with conserved variables the coordinate-frame density $D = \Gamma\rho$, the total energy density $E = \Gamma^2 w_{\rm gas} - P$, and the momentum density components $M^i = \Gamma^2 w_{\rm gas} v^i$. The primitive variables are the comoving rest-mass density $\rho$, gas pressure $P$, and the spatial components $u^i$ of the four-velocity, from which the code constructs the Lorentz factor $\Gamma = (1 + u^i u_i)^{1/2}$, the three-velocity $v^i = u^i/\Gamma$, and the comoving gas enthalpy density
\begin{equation}
w= w_{\rm gas} = \rho + \frac{\Gamma_{\rm adi}}{\Gamma_{\rm adi} - 1}\, P.
\label{eq:enthalpy}
\end{equation}
We use $\Gamma_{\rm ad} = 4/3$ and $c=1$ throughout. For the coldest test cases, where $p_0/\rho_0\ll1$, a non-relativistic ideal gas would more naturally be described by $\Gamma_{\rm ad}=5/3$. We adopt $\Gamma_{\rm ad}=4/3$ in these tests so that the same equation of state is used consistently in both the analytic calculation and the numerical simulations.

The radiation four-force $G^\nu$ on the right-hand side is applied as a user source term in its full nonlinear form, evaluated from the local primitive state at every step of the time integrator.
\begin{equation}
G^x = \frac{\sigma_{\rm T}}{m_e}\,\rho\, U'
\left[1 + \tfrac{2}{3}\langle\gamma^2\beta^2\rangle\right]
\end{equation}
with the comoving radiation energy density evaluated from the local velocity, $U' = \Gamma^2(1-\beta)^2 U_{\rm rad}$, and the Compton factor evaluated from the local density through the relativistic electron adiabat,
$\langle\gamma^2\beta^2\rangle = \langle\gamma^2\beta^2\rangle_0
(\rho/\rho_0)^{2/3}$.

The drag is parameterized at runtime by the inverse-Compton force fraction
\begin{equation}
\xi \equiv \frac{2\langle\gamma^2\beta^2\rangle_0}
{3 + 2\langle\gamma^2\beta^2\rangle_0} \in [0,1) ,
\end{equation}
together with the ambient radiation energy density $U_{\rm rad}$; the code inverts this relation to obtain $\langle\gamma^2\beta^2\rangle_0$. The characteristic drag frequency and sound speed that appear in the dispersion relation follow as $\omega_0 = f_0/w_0$ and $a_s^2 = \Gamma_{\rm ad}P_0/w_0$, with $f_0 = (\sigma_{\rm T}U_{\rm rad}/m_e)\rho_0 [1+\tfrac{2}{3}\langle\gamma^2\beta^2\rangle_0]$.

The domain is one-dimensional and Cartesian with periodic boundaries. For a box of
length $L$ and mode number $n_{\rm mode}$ the seeded wavenumber is $k = 2\pi n_{\rm mode}/L$; unless stated otherwise we use $L = 8\pi$ with $n_{\rm mode} = 4$, giving $k = 1$. A single eigenmode of the dispersion relation from Eq.~(\ref{eq:dispersion_relation})
is seeded as the initial condition, with a runtime flag selecting which of the two
complex roots is initialized. The density perturbation is $\delta\rho = A\sin(kx)$, the pressure follows the adiabatic closure $\delta p_{\rm gas} = (\Gamma_{\rm ad}P_0/\rho_0)\,\delta\rho$, and the velocity perturbation is constructed from the eigenvector ratio implied by the linearized continuity equation,
\begin{equation}
\frac{\delta\beta}{\delta\rho} =
\frac{i\,\omega}{\rho_0\,(ik + \omega_0)} \equiv r,
\end{equation}
as $\delta u^x = A\left[\mathrm{Re}(r)\sin(kx) + \mathrm{Im}(r)\cos(kx)\right]$.
The background is initially at rest, $\beta_0 = 0$.

The runs use the HLLC Riemann solver \citep{beckwith_second-order_2011}, piecewise-linear reconstruction, the second-order van Leer predictor-corrector integrator, and a CFL number of $0.3$. The standard $k=1$ runs use a periodic domain of length $L=8\pi$ with $n_{\rm mode}=4$, resolved with $N_x=4096$ cells, corresponding to $1024$ cells per wavelength. We adopt code units with $c=1$, $\sigma_{\rm T}=1$, $m_e=1$, $\rho_0=1$, varying $P_0$ and $\xi$, and initialize a density perturbation with amplitude $\delta\rho/\rho_0=10^{-3}$. Simulation snapshots are written in double precision for the subsequent mode analysis.

We consider four parameter sets, listed in Table~\ref{tab:wave_cases}, chosen to sample both the pressure and $\xi$ dependence of the dispersion relation. Tests~1 and~2 have $P_0/\rho_0=1.5\times10^{-2}$ and differ in $\xi$, while Tests~3 and~4 provide the corresponding pair at $P_0/\rho_0=10$. Equivalently, Tests~1 and~3 share the high-$\xi$ value, while Tests~2 and~4 share the low-$\xi$ value. For each parameter set, $U_{\rm rad}$ is chosen so that the characteristic drag frequency $\omega_0=f_0/w_0$ is the same. Each standard case is run twice, initializing the $\omega_+$ and $\omega_-$ eigenvectors separately.

In addition to the standard $k/\omega_0\simeq7850$ runs, we perform several simulations at smaller wavenumber to test the wavelength dependence of the dispersion relation. Test~1 is additionally evolved at $k/\omega_0=30$ and $0.1$, while Test~4 is evolved at $k/\omega_0=0.1$ to probe the long-wavelength growing branch.

\begin{table*}
  \centering
  \caption{Parameters of the four validation cases and the wavelengths used in the numerical tests. 
  Only $P_0/\rho_0$ and $\xi$ are varied between cases; $U_{\rm rad}$ is then chosen so that all four cases have the same characteristic drag frequency $\omega_0=f_0/w_0$. 
  Tests~1 and~2 share the same $P_0/\rho_0$ and differ in $\xi$, while Tests~3 and~4 provide the corresponding comparison at high $P_0/\rho_0$. 
  Each standard case is run at $k/\omega_0=7850$ for both the $\omega_+$ and $\omega_-$ eigenmodes. 
  Test~1 is additionally run at $k/\omega_0=30$ and $0.1$, and Test~4 at $k/\omega_0=0.1$, to probe the wavelength dependence of the unstable branches.}
  \label{tab:wave_cases}

  \begin{tabular}{lcccc}
    \toprule
     & Test 1 & Test 2 & Test 3 & Test 4 \\
    \hline

    \multicolumn{5}{l}{\emph{Input parameters}} \\

    $P_0/\rho_0$
        & $1.5\times10^{-2}$
        & $1.5\times10^{-2}$
        & $10$
        & $10$ \\

    $\xi$
        & $0.99975$
        & $0.25$
        & $0.99975$
        & $0.25$ \\

    $U_{\rm rad}$
        & $3.33\times10^{-8}$
        & $1.01\times10^{-4}$
        & $1.29\times10^{-6}$
        & $3.92\times10^{-3}$ \\

    \hline
    \multicolumn{5}{l}{\emph{Derived quantities}} \\

    $\langle\gamma^2\beta^2\rangle_0$
        & $6075$
        & $0.5$
        & $6075$
        & $0.5$ \\

    $w_0$
        & $1.06$
        & $1.06$
        & $41.0$
        & $41.0$ \\

    $a_s^2$
        & $0.019$
        & $0.019$
        & $0.325$
        & $0.325$ \\

    $\omega_0$
        & $1.274\times10^{-4}$
        & $1.274\times10^{-4}$
        & $1.274\times10^{-4}$
        & $1.274\times10^{-4}$ \\

    $\xi_{\rm low}$
        & $0.0283$
        & $0.0283$
        & $0.4878$
        & $0.4878$ \\

    $\xi_{\rm high}$
        & $0.4687$
        & $0.4687$
        & $2.6864$
        & $2.6864$ \\

    $\xi_{\rm min}$
        & $0.0291$
        & $0.0291$
        & $0.9524$
        & $0.9524$ \\

    stability regime
        & growing
        & damped
        & damped
        & growing \\

    &
        ($\xi>\xi_{\rm high}$)
        &
        ($\xi_{\rm low}<\xi<\xi_{\rm high}$)
        &
        ($\xi_{\rm low}<\xi<\xi_{\rm high}$)
        &
        ($\xi<\xi_{\rm low}$) \\

    \hline
    \multicolumn{5}{l}{\emph{Simulation runs}} \\

    $k/\omega_0 = 7850$
        & $\omega_+,\ \omega_-$
        & $\omega_+,\ \omega_-$
        & $\omega_+,\ \omega_-$
        & $\omega_+,\ \omega_-$ \\

    $k/\omega_0 = 30$
        & $\omega_+,\ \omega_-$
        & --
        & --
        & -- \\

    $k/\omega_0 = 0.1$
        & $\omega_+,\ \omega_-$
        & --
        & --
        & $\omega_+,\ \omega_-$ \\

    \hline
  \end{tabular}
\end{table*}

\subsection{\label{sec:analytic_numeric_comparison}Analytic-numerical comparison}
To validate the analytic dispersion relation, we measure the complex mode frequency $\omega$ directly from the simulations. For a perturbation evolving as
\begin{equation}
\delta \rho_k(t) \propto \exp[-i\omega t],
\end{equation}
with $\omega=\mathrm{Re}(\omega)+i\,\mathrm{Im}(\omega)$, the mode amplitude evolves as
\begin{equation}
\ln |\delta \rho_k(t)| = \ln |\delta \rho_k(0)|
+ \mathrm{Im}(\omega)t,
\end{equation}
while its phase evolves as
\begin{equation}
\phi_k(t) = \phi_k(0) - \mathrm{Re}(\omega)t.
\end{equation}
Thus, the growth or damping rate $\mathrm{Im}(\omega)$ is obtained from the slope of the logarithmic mode amplitude, while $\mathrm{Re}(\omega)$ is obtained from the phase evolution.

We measure the wave amplitude using the root-mean-square density perturbation over the full computational domain,
\begin{equation}
    A(t) = \sqrt{2}\,\langle\delta\rho^2\rangle^{1/2}.
\end{equation}
For a sinusoidal perturbation $\delta\rho=A\cos(kx+\phi)$, $\langle\delta\rho^2\rangle=A^2/2$, so the factor of $\sqrt{2}$ recovers the physical wave amplitude. This measure is also independent of the instantaneous phase of the propagating wave. Since the change in amplitude between successive snapshots is small, averaging over all grid cells reduces sensitivity to numerical noise, grid-scale fluctuations, and the motion of the wave relative to the mesh. The measured amplitude therefore follows
\begin{equation}
    \ln A(t)
    =
    \ln A(0)
    +
    \mathrm{Im}(\omega)t,
\end{equation}
from which $\mathrm{Im}(\omega)$ is obtained by a linear fit.

%%%%%%%%%%%%%%%%%%%%%%%%%%%%%%%%%%%%%%%%%
The four simulation tests were chosen to sample different regions of the parameter space shown in Fig.~\ref{fig:max_rate}. Test 1 lies in the growing region, while Tests 2 and 3 lie in the damped region. For each parameter set, we initialize both roots of Eq.~\ref{eq:dispersion_roots} at one fixed frequency, giving eight independent linear-wave tests. The simulations confirm that the Test 1 $\omega_+$ mode grows, Test 4 $\omega_-$ mode grows, while the other six modes decay.

\begin{figure}
    \begin{center}  
    \includegraphics[width=\linewidth]{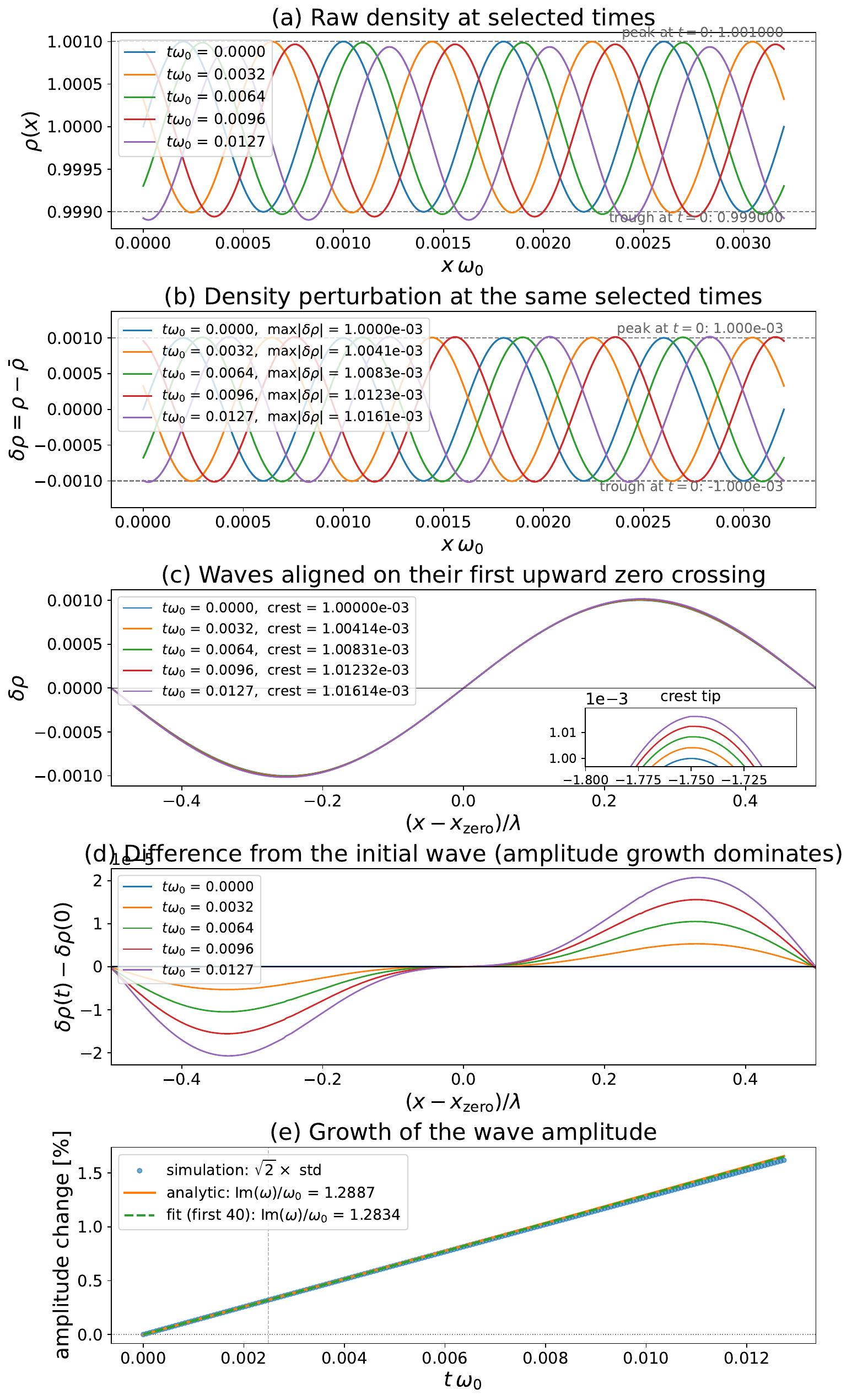}
    \end{center}
    \caption{Linear-wave validation for Test 1, growing mode, over a short-time interval $t\omega_0\in[0,0.1]$ where the local fixed-background approximation is good. Five snapshots evenly spaced over time are shown. (a) Raw density profile at each snapshot time. (b) The corresponding density perturbation $\delta\rho=\rho-\bar\rho$. (c) The same curves shifted so that each snapshot's first upward zero crossing lies at the origin, removing the phase (horizontal) motion; the inset zooms on the crest to show the small amplitude increase directly. (d) Each snapshot's perturbation minus the initial ($t\omega_0=0$) profile, isolating the growth from the underlying wave shape. (e) Wave amplitude, $A=\sqrt{2}\langle\delta\rho^2\rangle^{1/2}$, versus time: simulated values (points) compared against the analytic growth rate $\mathrm{Im}(\omega)/\omega_0=1.2887$ (solid line) and an exponential fit to the simulation $\mathrm{Im}(\omega)/\omega_0=1.2741$ (dashed line).}
    \label{fig:Test1_linear_validation}
\end{figure}

We first illustrate the evolution of the growing Test 1 $\omega_+$ mode in Fig.~\ref{fig:Test1_linear_validation}. The growing eigenmode propagates while maintaining the same sinusoidal shape over a short time interval. In the raw profiles in panel (a), the perturbation wave moves to the right while the amplitude increases, with a small downward drift of the mean density which arises from the Lorentz factor $\gamma \leq 1$ in the relation $\rho = D/\gamma$ between the conserved mass density $D$ and the primitive density $\rho$. After removing the mean drift by centering the curves vertically in panel (b), and removing the horizontal phase motion by aligning the first upward zero crossing of each snapshot, we can see the amplitude growth directly from panel (c), where the small inset inside zooms in at the crest to show the small increase from one snapshot to the next. Panel (d) shows each selected wave subtracting the initial wave profile, isolating the growth of the perturbation. The mode remains the same sinusoidal form with increasing amplitude. The slight distortion in the middle is from the second harmonics appearing in the nonlinear terms $\propto \delta \rho \delta \beta \propto \sin(2kx)$, which will be stronger when we run longer times and will be discussed in the Caveats section. The amplitude of the second harmonic is very small (0.01\% over the fitting range) and within our relative error, thus does not affect the results. And last, panel (e) shows a direct numerical validation of the analytic growth rate for this case.

\begin{figure}
    \begin{center}  
    \includegraphics[width=\linewidth]{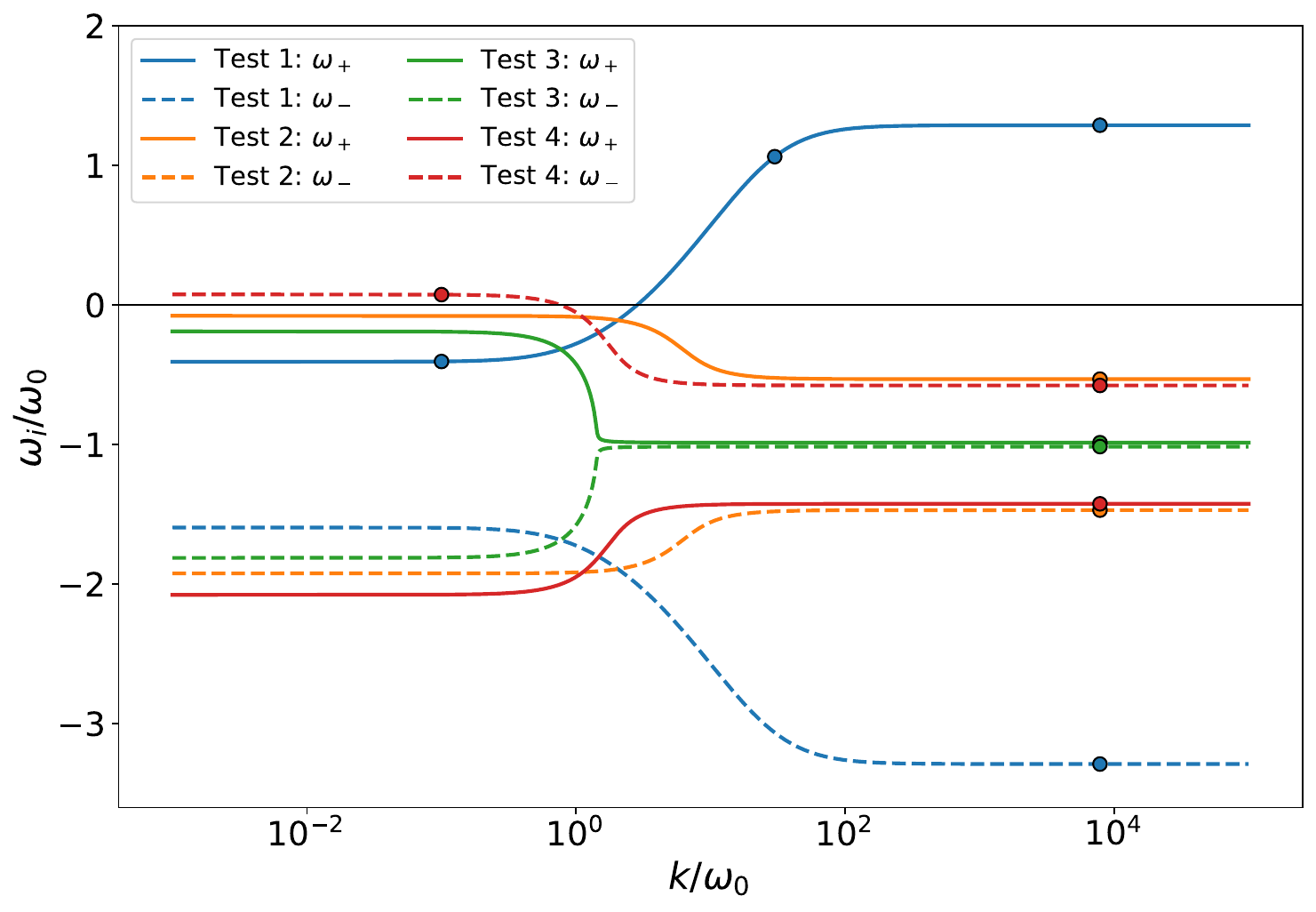}
    \end{center}
    \caption{Normalized growth and damping rates $\omega_i/\omega_0$ as a function of $k/\omega_0$ for the four test cases, from the two roots [Eq.~(\ref{eq:dispersion_roots})] of the dispersion relation in Eq.~(\ref{eq:dispersion_relation}). Solid curves are the $\omega_{+}$ root, dashed curves the $\omega_{-}$ root, and the circles mark the simulated wavenumber $k/\omega_0 = 7850$ (k=1). Test 1 has a positive root at large k in $\omega_{+}$ mode, while Test 4 has a positive root at small k in $\omega_{-}$ mode.}
    \label{fig:three_case_dispersion}
\end{figure}

We next compare the full wavenumber dependence of the analytic dispersion relation for the four test cases in Fig.~\ref{fig:three_case_dispersion}.
Test~1 has a positive $\omega_+$ branch at large $k/\omega_0$, whereas its $\omega_-$ branch remains damped. Both branches of Tests~2 and~3 remain damped over the wavenumber range shown. Test 4 has a positive $\omega_-$ branch at small $k/\omega_0$, while its $\omega_+$ branch remains damped. We further test the wavenumber dependence of Eq.~(\ref{eq:dispersion_roots}) by repeating the Test~1 $\omega_+$ run at $k/\omega_0 = 30$ and $0.1$, and repeating the Test~4 $\omega_-$ run and at $k/\omega_0 = 0.1$, holding all other parameters fixed. The fiducial runs sit at $k/\omega_0 = 7850$ on the short-wavelength plateau, the additional runs probe the transition and the long(short)-wavelength limit. For Test 1, the same mode that grows at $k/\omega_0 = 7850$ and $30$ is damped at $k/\omega_0 = 0.1$, since short-wavelength growth instead requires $\xi > \xi_{\rm high}$. For Test 4, the same mode that dampens at $k/\omega_0 = 7850$ is unstable at $k/\omega_0 = 0.1$, since long-wavelength growth instead requires $\xi < \xi_{\rm low}$. 

\begin{figure}
    \begin{center}  
    \includegraphics[width=\linewidth]{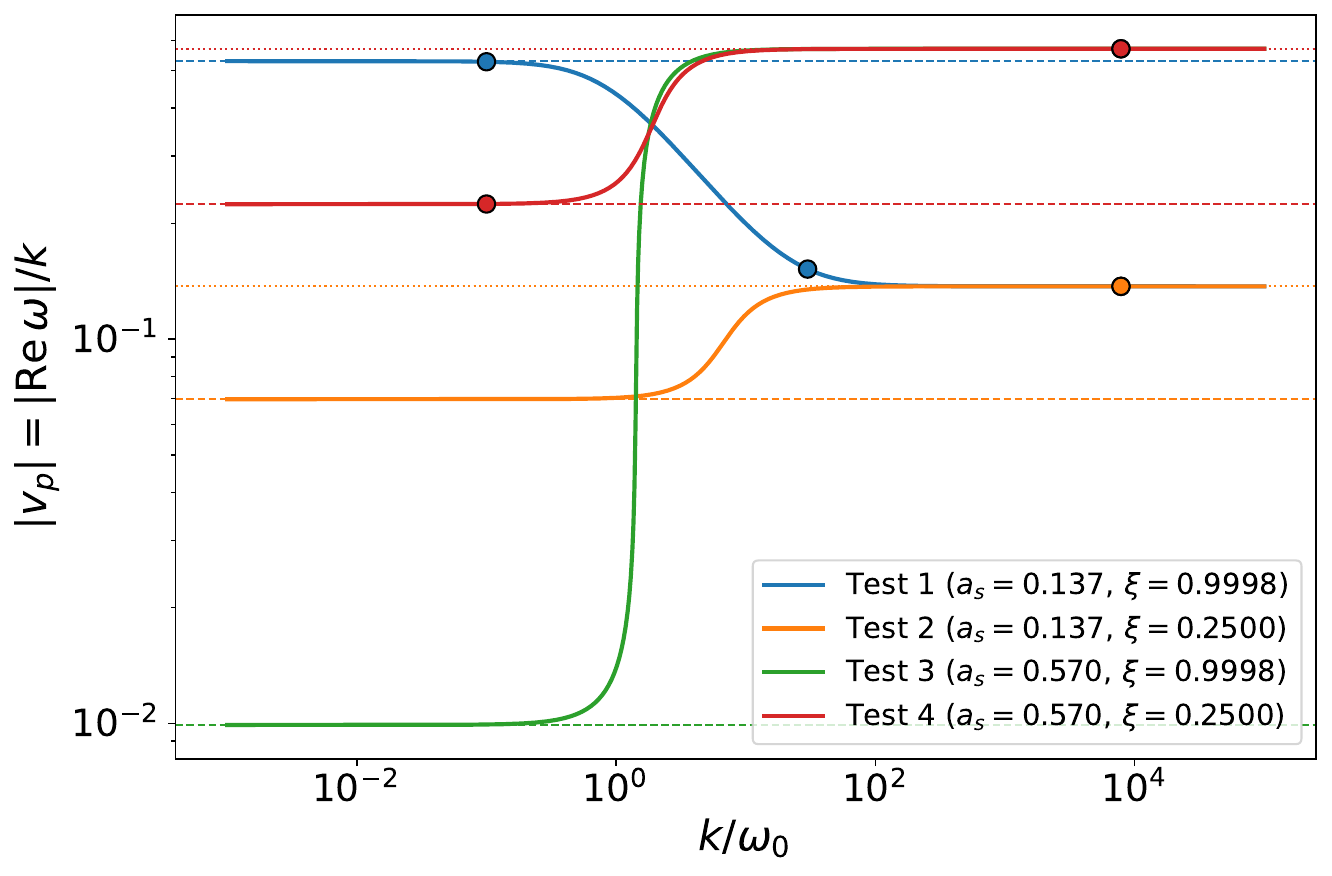}
    \end{center}
    \caption{Phase velocity $v_p=\mathrm{Re}(\omega)/k$ as a function of $k/\omega_0$ for the four test cases. The wavenumbers of the simulated modes are marked on each curve. At large $k/\omega_0$ (right), the phase velocity approaches the ordinary sound speed, $v_p\to a_sc$ (horizontal dotted lines), independent of $\xi$. Tests~1 and~2 share $a_s=0.137$ and converge to the same plateau despite their very different $\xi$, while Test~3 and 4's larger $a_s=0.570$ places them on a higher plateau. At small $k/\omega_0$ (left), radiation drag dominates the dynamics and $v_p$ instead depends jointly on $\xi$ and $a_s$. }
    \label{fig:three_case_phase_velocity}
\end{figure}
The corresponding phase velocities are shown in Fig.~\ref{fig:three_case_phase_velocity}.
When $k/\omega_0\gg1$, the oscillation timescale is short compared with the radiative-drag timescale and $v_p\rightarrow a_s c$. The influence of the radiation force becomes more apparent toward small $k/\omega_0$. For $k<<\omega_0$, 
\begin{equation}
    v_p \approx \frac{|\frac{\xi}{3}-a_s^2|}{\sqrt{\left|\frac{2\xi}{3}-1-a_s^2\right|}},
    \label{eq:vpask0}
\end{equation}
each wave goes to a different constant depending on the values of $\xi$ and $a_s$ as shown.

\begin{figure}
    \begin{center}  
    \includegraphics[width=\linewidth]{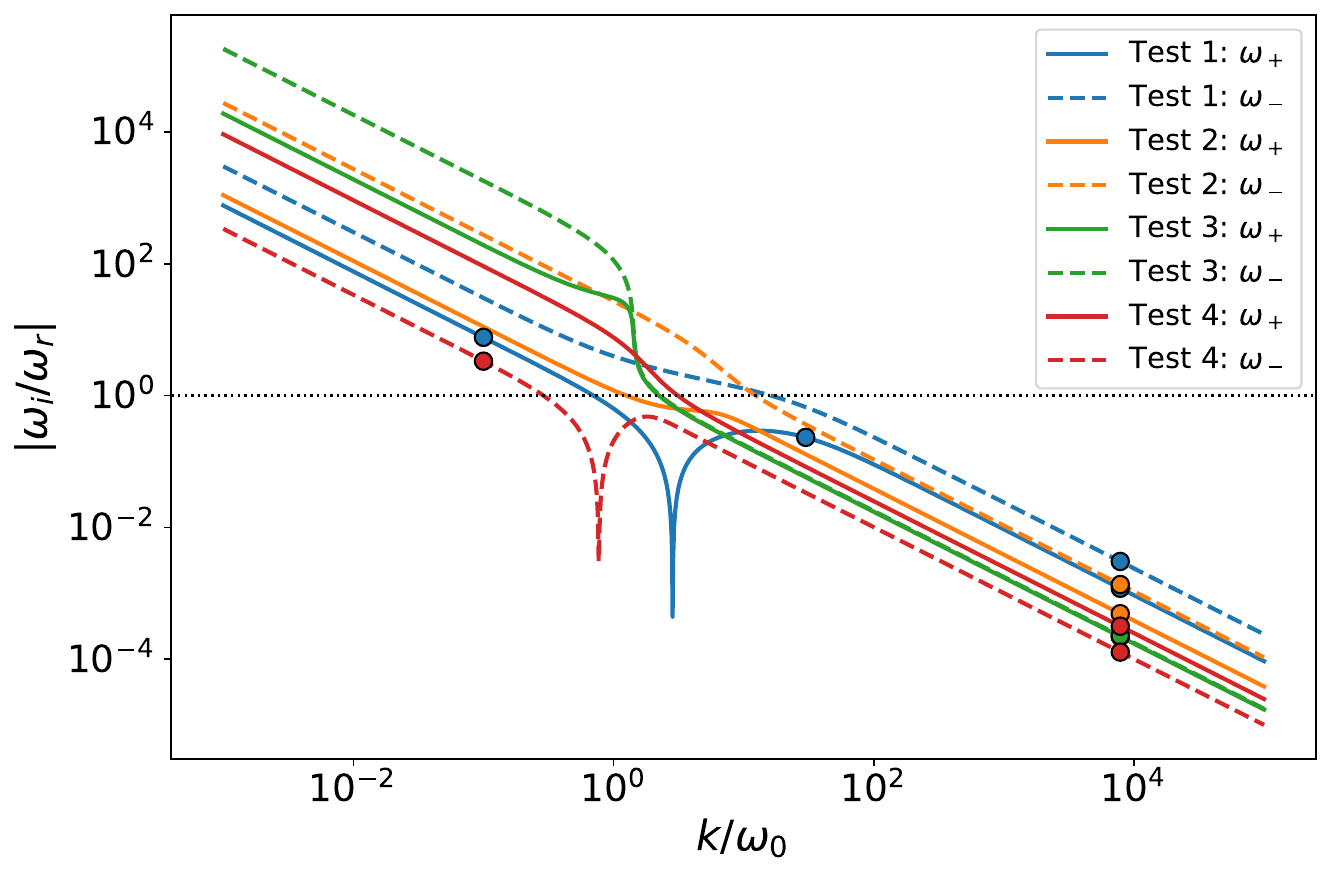}
    \end{center}
    \caption{Ratio of the imaginary to real part of the frequency, $|\omega_i/\omega_r|$, as a function of $k/\omega_0$ for the two roots ($\omega_+$, solid; $\omega_-$, dashed) of each of the four test cases. The wavenumbers of the simulated modes are marked on each curve. The horizontal dotted line marks $|\omega_i/\omega_r|=1$. The sharp downward spike unique to Test~1's $\omega_+$ curve, near $k/\omega_0\sim2$, marks the wavenumber at which that branch's growth rate changes sign from damped to growing (likewise for Test 4), so $|\omega_i/\omega_r|\to0$ exactly at the crossing; Tests~2 and~3 remain damped at all $k$ and never show this feature.}
    \label{fig:three_case_ratio}
\end{figure}

The relative importance of the imaginary and real parts of the mode
frequency is shown in Fig.~\ref{fig:three_case_ratio}. At large $k/\omega_0$, $|\omega_i/\omega_r|\ll1$, so many wave oscillations occur over one growth or damping time. Toward longer wavelengths the two timescales become comparable. The zero in the Test~1 $\omega_+$ curve marks the transition from a damped mode at small $k/\omega_0$ to the growing short-wavelength branch sampled by the fiducial simulation.

We quantify the agreement between the simulated and analytic mode frequencies in Table~\ref{tab:wave_test_results}.

\begin{table*}
  \centering
  \caption{Measured mode frequencies compared with the analytic dispersion relation.
  The standard validation runs use $k=1$ ($k/\omega_0 \simeq 7850$) for all four parameter sets in Table~\ref{tab:wave_cases}. Additional runs for Tests~1 and~4 probe the wavelength dependence of the dispersion relation. All values are extracted from over $0\le \omega_0 t\le 2.548\times 10^{-3}$ (dimensionless time).
   A straight-line fit $\varphi(t)=\varphi_0-\mathrm{Re}\,\omega\,t$ gives $\mathrm{Re}\,\omega$ as minus the slope.
  $\mathrm{Im}\,\omega$ is obtained from the slope of $\ln A(t)$, with
  $A=\sqrt{2}\,\langle\delta\rho^2\rangle^{1/2}$.}
  \label{tab:wave_test_results}

  \begin{tabular}{llccc}
    \toprule
    Run & Quantity & Analytic & Measured & Rel.\ err. \\
    \hline

    \multicolumn{5}{l}{\emph{Standard runs: $k=1$ ($k/\omega_0 \simeq 7850$)}} \\

    Test 1, $\omega_+$ (growing)
      & $\mathrm{Re}\,\omega$ & $+1.373\times10^{-1}$ & $+1.386\times10^{-1}$ & $0.89\%$ \\
      & $\mathrm{Im}\,\omega$ & $+1.642\times10^{-4}$ & $+1.635\times10^{-4}$   & $0.42\%$ \\

    Test 1, $\omega_-$ (damped)
      & $\mathrm{Re}\,\omega$ & $-1.374\times10^{-1}$ & $-1.361\times10^{-1}$ & $0.89\%$ \\
      & $\mathrm{Im}\,\omega$ & $-4.190\times10^{-4}$ & $-4.183\times10^{-4}$   & $0.15\%$ \\

    \hline

    Test 2, $\omega_+$ (slow-damped)
      & $\mathrm{Re}\,\omega$ & $+1.374\times10^{-1}$ & $+1.386\times10^{-1}$ & $0.89\%$ \\
      & $\mathrm{Im}\,\omega$ & $-6.760\times10^{-5}$ & $-6.769\times10^{-5}$   & $0.13\%$ \\

    Test 2, $\omega_-$ (fast-damped)
      & $\mathrm{Re}\,\omega$ & $-1.374\times10^{-1}$ & $-1.361\times10^{-1}$ & $0.89\%$ \\
      & $\mathrm{Im}\,\omega$ & $-1.872\times10^{-4}$ & $-1.870\times10^{-4}$   & $0.02\%$ \\

    \hline

    Test 3, $\omega_+$ (damped)
      & $\mathrm{Re}\,\omega$ & $+5.703\times10^{-1}$ & $+5.711\times10^{-1}$ & $0.15\%$ \\
      & $\mathrm{Im}\,\omega$ & $-1.256\times10^{-4}$ & $-1.255\times10^{-4}$   & $0.04\%$ \\

    Test 3, $\omega_-$ (damped)
      & $\mathrm{Re}\,\omega$ & $-5.703\times10^{-1}$ & $-5.669\times10^{-1}$ & $0.15\%$ \\
      & $\mathrm{Im}\,\omega$ & $-1.292\times10^{-4}$ & $-1.293\times10^{-4}$   & $0.12\%$ \\

    \hline

    Test 4, $\omega_+$ (damped)
      & $\mathrm{Re}\,\omega$ & $+5.703\times10^{-1}$ & $+5.711\times10^{-1}$ & $0.15\%$ \\
      & $\mathrm{Im}\,\omega$ & $-1.814\times10^{-4}$ & $-1.812\times10^{-4}$   & $0.14\%$ \\

    Test 4, $\omega_-$ (damped)
      & $\mathrm{Re}\,\omega$ & $-5.703\times10^{-1}$ & $-5.694\times10^{-1}$ & $0.14\%$ \\
      & $\mathrm{Im}\,\omega$ & $-7.336\times10^{-5}$ & $-7.352\times10^{-5}$   & $0.23\%$ \\

    \hline
    \multicolumn{5}{l}{\emph{Additional wavelength tests}} \\

    Test 1, $\omega_+$, $k/\omega_0=30$ (growing)
      & $\mathrm{Re}\,\omega$ & $+5.822\times10^{-4}$ & $+5.859\times10^{-4}$ & $0.63\%$ \\
      & $\mathrm{Im}\,\omega$ & $+1.355\times10^{-4}$ & $+1.354\times10^{-4}$   & $0.09\%$ \\

    Test 1, $\omega_+$, $k/\omega_0=0.1$ (damped)
      & $\mathrm{Re}\,\omega$ & $+6.722\times10^{-6}$ & $+6.773\times10^{-6}$ & $0.11\%$ \\
      & $\mathrm{Im}\,\omega$ & $-5.149\times10^{-5}$ & $-5.150\times10^{-5}$   & $0.01\%$ \\

    \hline

    Test 4, $\omega_-$, $k/\omega_0=0.1$ (growing)
      & $\mathrm{Re}\,\omega$ & $-2.866\times10^{-6}$ & $-2.847\times10^{-6}$ & $0.65\%$ \\
      & $\mathrm{Im}\,\omega$ & $+9.564\times10^{-6}$ & $+9.397\times10^{-6}$ & $1.75\%$ \\

    \hline
  \end{tabular}
\end{table*}

For the eleven fiducial modes, the measured real frequencies agree with the analytic values to within $0.02\%$--$1.75\%$. The agreement holds for both signs of $\mathrm{Re}\,\omega$ and for both growing and damped solutions $\mathrm{Im}\,\omega$.

\section{\label{sec:discussion}Discussion}
\subsection{Large and small wavelength limits of the dispersion relation}
To understand the larger separations among the tests at long wavelengths and smaller separations at short wavelengths in Fig.~\ref{fig:three_case_dispersion}, we look at the asymptotic limits directly.

For the short-wavelength limit $k/\omega_0\to\infty$, the
real part of $\omega$ approaches
\begin{equation}
\frac{\mathrm{Re}(\omega)}{\omega_0} \to \pm a_s\frac{k}{\omega_0} \quad\Longrightarrow\quad v_p \to \pm a_s c ,
\end{equation}
and the imaginary part in Eq.~\ref{eq:omega_i_at_k_infty} approaches
\begin{equation}
\frac{\omega_i}{\omega_0} \to -1 \pm a_s\left(1-\frac{\xi}{3a_s^2}\right).
\end{equation}
The short-wavelength phase velocity depends on $a_s$ alone, recovering the relativistic sound speed, but the mode remains either growing or damped with a growth/damping rate that still depends on $\xi$ and $a_s$, so the wave is not a pure acoustic wave.

For $k/\omega_0\to0$, $\mathrm{Re}(\omega)\propto k$ $\to0$ reaches a constant at fixed $\xi$ and $a_s$ as in Eq.~\ref{eq:vpask0}
\begin{equation}
    \mathrm{Re}(\omega) = \omega_0 k \frac{|\frac{\xi}{3}-a_s^2|}{\sqrt{\frac{2\xi}{3}-1-a_s^2}}
\end{equation}
and in Eq.~\ref{eq:omega_i_at_k_0}
\begin{equation}
\frac{\omega_i}{\omega_0} \to -1 \pm \sqrt{1+a_s^2-\frac{2\xi}{3}} .
\end{equation}
At long wavelength the radiation force dominates the dynamics faster than pressure restoring forces can act, so the mode is fundamentally a radiation-drag mode rather than a weakly perturbed sound wave, and its growth/damping rate is again set jointly by $\xi$ and $a_s$.

Evaluating the large-$k$ limit for each test case,
\begin{equation}
\begin{aligned}
\text{Test 1:}\quad \frac{\omega_i}{\omega_0}\Big|_{k\to\infty} &= -1\pm2.296 \to \{+1.296,\ -3.296\}\\
\text{Test 2:}\quad \frac{\omega_i}{\omega_0}\Big|_{k\to\infty} &= -1\pm0.471 \to \{-0.529,\ -1.471\}\\
\text{Test 3:}\quad \frac{\omega_i}{\omega_0}\Big|_{k\to\infty} &= -1\pm0.0147 \to \{-0.985,\ -1.015\}\\
\text{Test 4:}\quad \frac{\omega_i}{\omega_0}\Big|_{k\to\infty} &= -1\pm0.424 \to \{-0.576,\ -1.424\}
\end{aligned}
\end{equation}
matches the plateaus on the right-hand side of Fig.~\ref{fig:three_case_dispersion}.

Evaluating the small-$k$ limit,
\begin{equation}
\begin{aligned}
\text{Test 1:}\quad \frac{\omega_i}{\omega_0}\Big|_{k\to0} &= -1\pm0.594 \to \{-0.407,\ -1.594\}\\
\text{Test 2:}\quad \frac{\omega_i}{\omega_0}\Big|_{k\to0} &= -1\pm0.923 \to \{-0.077,\ -1.923\}\\
\text{Test 3:}\quad \frac{\omega_i}{\omega_0}\Big|_{k\to0} &= -1\pm0.811 \to \{-0.189,\ -1.811\}\\
\text{Test 4:}\quad \frac{\omega_i}{\omega_0}\Big|_{k\to0} &= -1\pm0.076 \to \{+0.076,\ -2.076\}
\end{aligned}
\end{equation}
matches the flat plateaus on the left-hand side of Fig.~\ref{fig:three_case_dispersion}.

Since $v_p\to a_sc$ at large $k$ depends only on $a_s$, Tests 1 and 2 which share $a_s=0.137$ but have very different $\xi$, converge onto the same phase-velocity plateau in Fig.~\ref{fig:three_case_phase_velocity} despite lying in opposite stability regimes. Similarly, Test 3 and 4 share $a_s=0.570$ and converge to the same higher plateau. The large-k plateau is thus set entirely by $a_s$, while the small-k region depends on both $\xi$ and $a_s$.

In Fig.~\ref{fig:three_case_ratio}, at large $k$ the real part grows linearly, $\omega_r\propto a_sk$, while $\omega_i$ approaches a constant, so $|\omega_i/\omega_r|\propto1/k$ a downward-sloping line on log-log axes, seen for all six curves. At small $k$, $\omega_r$ is itself linear in $k$ while $\omega_i$ approaches a different constant, so $|\omega_i/\omega_r|\propto1/k$ diverges as $k\to0$ as well, producing the steep rise on the left of the figure. The sharp notch unique to Test~1's $\omega_+$ curve arises because $\omega_i$ for that branch varies continuously from $-0.407$ at small $k$ to $+1.296$ at large $k$, and must therefore pass through exactly zero at some intermediate wavenumber $k^*$ (same for Test~4 $\omega_-$) at that point $|\omega_i/\omega_r|\to0$, producing a sharp downward spike rather than a smooth minimum. Tests 2 and 3 never change sign and therefore never produce this feature.

\subsection{Physical interpretation: which terms drive the damping/growth?}
We can interpret the growth and damping directly from the momentum equation Eq.~\ref{eq:perturbedmomentum}
\begin{align}
    \underbrace{w_0\frac{\partial\delta\beta}{\partial t}}_{\text{acceleration}}
 & + \underbrace{\left(1 + \frac{\Gamma_{\rm ad}}{\Gamma_{\rm ad}-1}\frac{P_{0}}{\rho_{0}}\right)\frac{f_{0}}{w_{0}}\delta\rho}_{\text{inertial}}
+ \underbrace{\frac{\partial\delta P}{\partial x}}_{\text{acoustic}}
= \underbrace{\delta f^1}_{\text{force}} ,
\nonumber
\end{align}
where the direct radiation-force perturbation is
\begin{equation}
\delta f^1 = f_0\left[\underbrace{-2\,\delta\beta}_{\text{velocity-dep drag}} + \underbrace{\left(1+\tfrac{2\xi}{3}\right)\frac{\delta\rho}{\rho_0}}_{\text{density-dep drag}}\right] .
\nonumber
\end{equation}
Whether a mode grows or damps depends on the sum of these terms.
The relative phase relation of $\delta\rho$ and $\delta\beta$ is set by the continuity equation, $\delta\rho/(\rho_0\delta\beta)=(ik+\omega_0)/(i\omega)$. %A right-moving mode has $\delta \rho$ and $\delta \beta$ in phase and a left-moving mode has them out of phase.

The \emph{acceleration} term, $w_0\,\partial_t\delta\beta$ on the left hand side encodes the evolution of a velocity perturbation. Naturally, the overall acceleration is reduced for large values of the mean fluid inertia $w_{0}$. We can evaluate whether a perturbation is damped or grows from the sign of this term, which we can evaluate from all the other terms.

The \emph{inertial response to the background acceleration},
$\left(\delta\rho + \tfrac{\Gamma_{\rm ad}}{\Gamma_{\rm ad}-1}\delta P\right)\partial_t\beta_0
= \delta w\,\partial_t\beta_0$, arises because the background flow is being
accelerated by the radiation at the rate $\partial_t\beta_0=f_0/w_0$. A region with
increased enthalpy $\delta w > 0$ has more inertia than the mean flow, which reduces the acceleration, and vice versa.

The \emph{pressure gradient}, $\partial_x\delta P$, is the ordinary acoustic restoring force.

On the right hand side, the \emph{radiation force term} $\delta f^1$ has two parts. The density-dependent part, $f_0(1+\tfrac{2\xi}{3})\,\delta\rho/\rho_0$, is stronger where the gas is denser, as more particles lead to increased drag The velocity-dependent part, $-2f_0\,\delta\beta$, derives
from $\delta U'=-2\,\delta\beta\,U_0'$. It implies that fluid moving faster (larger value of $\beta$) has a lower velocity relative to the radiation field, and the fluid will experience a lower drag force due to reduced Doppler boosting. 

In the following, we distinguish right- and left-propagating modes, i.e., modes with $\omega_{\rm r,+} > 0$ and $\omega_{\rm r,-} < 0$, respectively.  These correspond to co- and counter-propagating modes, respectively, with respect to the direction of the radiation drag force in the fluid, which is in the positive x-direction as defined above.

\subsubsection*{Long-wavelength modes}
At long wavelengths, the density gradient is shallow, so the pressure term $\partial_x\delta P$ is negligible, and the force balance is between the inertial term, the density-dependent force, and the velocity-dependent force. These modes propagate very slowly. As can be seen from Fig.~\ref{fig:three_case_ratio}, in the small-k regime, $\omega_{\rm i}\gg \omega_{r}$ and all modes are in the over-damped regime, that is, the flow does not undergo any oscillations and any perturbations grow or damp in place. The phase and group velocity differ from the sound speed. Since the growth/damping times are also longer than the acceleration time, our assumption of a non-relativistic fluid rest frame breaks down before the fluid can be materially affected, and a fully-relativistic treatment (including any effects of causality and simultaneity) must be undertaken to understand the long-term evolution of these modes.

We can understand these modes simply as the consequence of non-uniform fluids of different enthalpy and electron density (and thus drag force) experiencing different rates of acceleration. In this sense, the modes are physically of limited novelty: the growth or damping simply reflects the behavior of a non-uniform fluid subject to radiation drag. In the case of the derivation above, the non-uniformity is limited to adiabatic perturbations, which restrict the response of the fluid to the solutions of the dispersion relation.

\subsubsection*{Short-wavelength modes}
The short wavelength modes are physically more interesting. As we can see from Fig.~\ref{fig:three_case_ratio}, these modes are under-damped and undergo many oscillations within a growth/damping time and within an acceleration time. They behave like modified acoustic waves, either damped or unstable. At short wavelength the pressure gradient becomes important, leading to quasi-acoustic oscillations with phase velocity  very close to  the sound speed (see Fig.~\ref{fig:three_case_phase_velocity}).

Because all large-k modes are under-damped, we can assume that velocity and density perturbations are approximately in phase for the co-propagating mode and approximately 180 degrees out of phase for the counter-propagating mode. This is born out by the simulations and Eq.~(\ref{eq:deltarho_deltabeta_relation}) in the $k\gg\omega_0$ limit.

For both modes, the velocity-dependent force term is:
\begin{equation}
    f_\beta=-2f_0\delta\beta
\end{equation}
Because this term is negative, it is always damping: If the velocity perturbation is {\em positive}, the fluid velocity relative to the radiation field is {\em reduced} and the fluid experiences {\em less} drag force, i.e., {\em less} acceleration, {\em reducing} the velocity amplitude and thus damping the perturbation, while a {\em negative} velocity perturbation experiences {\em more} drag, {\em increasing} the velocity, and thus {\em also} reducing/damping the perturbation.

The other radiation-force-related terms depend on the density perturbation. First, consider the density-dependent drag force term:
\begin{equation}
    f_{\rho}=f_{0}\left(1 + \frac{2\xi}{3}\right)\frac{\delta \rho}{\rho_0}
\end{equation}
For the {\em co}-propagating mode (a velocity perturbation in the direction of the photon field), density and velocity are {\em in} phase. Consider a positive velocity perturbation. Then the density is also increased, leading to a {\em larger} radiation drag term, {\em accelerating} the fluid and thus contributing to {\em growth} of the velocity perturbation. 

The opposite is true for the {\em negative} density perturbation of the {\em counter}-propagating mode ($\omega_{-}$), for which this term is {\em damping}.

Finally, consider the inertial term:
\begin{equation}
    f_{w}=-f_0\left( 1 + \Gamma_{\rm ad}\frac{\Gamma_{\rm ad}}{\Gamma_{\rm ad} - 1}\frac{P_0}{w_0}\right)\frac{\delta\rho}{w_0}
\end{equation}
This term represents the fact that a perturbation that {\em increases} density and pressure adds inertia to the zero-order flow acceleration $\partial \beta_0/\partial t$ and thus requires a {\em larger} force to accelerate than the pure background flow. If not balanced by a corresponding {\em increase} in the radiative force $f_{\rho}$, it leads to a {\em decrease} in acceleration for a positive density/pressure perturbation and an {\em increase} in acceleration for negative density/pressure perturbation. As such it acts as a {\em damping} term for the {\em co-propagating} mode (where density and velocity are in phase) and as a {\em growth} term for the {\em counter-propagating} mode.

The free energy for any growth in this flow results from the anisotropy of the radiation field in the fluid rest frame (similar to a two-stream instability.)

Which term dominates (and thus whether a mode grows or is damped) depends on the sign of the net force (excluding the acoustic/pressure gradient term):
\begin{align}
    f_{\rm net} & = f_0 \left(\frac{2\xi}{3} - a_{\rm s}^2\right)\frac{\delta \rho}{\rho_0} - 2\delta\beta \\    
    & \sim f_0\left[\frac{2\xi}{3} - a_{\rm s}^2 \mp 2 a_{\rm s}\right] \frac{\delta\rho}{\rho_0} \\
    & \sim f_0\left[\pm\left(\frac{2\xi}{3} - a_{\rm s}^2\right) - 2a_{\rm s}\right] \frac{\delta\beta}{a_{\rm s}}
\end{align}
where the plus sign in the last row corresponds to the co-propagating (in-phase) mode and the minus sign to the counter-propagating (out-of-phase) mode and where we made use of the quasi-acoustic approximation $\delta\beta \sim \pm a_{s}\delta\rho/\rho_{0}$ from Eq.~(\ref{eq:deltarho_deltabeta_relation}). 

We distinguish the two limiting cases:
\begin{itemize}
\item{In the non-relativistic limit, $P_0\ll\rho_0$, we have $w_0\sim\rho_0$ and $a_{s}^2 \ll 1$ and this term becomes
\begin{equation}
    f_{\rm net}\sim f_0\left[\pm\frac{2\xi}{3} - 2a_{\rm s}\right] \frac{\delta\beta}{a_{\rm s}}
\end{equation}
The negative root corresponds to the counter-propagating mode, in which case both terms are {\em stabilizing/damping}, due to the out-of-phase nature of the density and velocity perturbations (see above).

The co-propagating (in-phase, $\omega_{+}$) mode can be {\em destabilized} if $\xi > 3a_{\rm s}$. This is the limit from eq.~(\ref{eq:large_k_limit}) for $a_{\rm s}\ll 1$. The physical mechanism at play here is the fact that the $\langle\gamma^2\rangle$ term grows faster than all other terms under adiabatic compression.}
\item{
In the relativistic limit $P_{0}\gg\rho_{0}$, we have $a_{\rm s}^2\sim 1/3$, so
\begin{align}    
    f_{\rm net} & \sim f_0\left[\pm\left(\frac{2\xi}{3} - \frac{1}{3}\right) - 2a_{\rm s}\right]\frac{\delta\beta}{a_{\rm s}} \\ & \leq f_0\left(1/3 - 2\sqrt{1/3}\right)\frac{\delta\beta}{a_{\rm s}}
\end{align}
as discussed above. Here, all modes are damped because the velocity-dependent term always dominates and the sign of the quantity in parentheses is always negative.}
\end{itemize}

%\subsubsection*{Instability in the non-relativistic limit}
%The relativistic term comes from the enthalpy $w_0 = \rho_0 +\frac{\Gamma_{\rm ad}}{\Gamma_{\rm ad}-1}P_0$ which has a pressure term adding to %the normal rest mass density. The background acceleration $\partial_t\beta_0$ acts on the enthalpy
%perturbation $\delta w=\delta\rho+\tfrac{\Gamma_{\rm ad}}{\Gamma_{\rm ad}-1}\delta P$. In the non-relativistic limit. the term would reduce to %$\delta\rho\,\partial_t\beta_0$.
%
%When $P0\ll\rho_0$, $a_s\rightarrow0$ the condition for the left moving mode to be unstable $\xi < \xi_{rm low} = \frac{3}{2}a_s^2$ is harder to %satisfy. The condition for the right moving mode to be unstable $\xi>\xi_{\rm high} = 3(a_s^2+a_s)$ also goes to 0, but $\xi$ by definition %needs to have relativistic electron motions $<\gamma^2 \beta^2>$. So the radiation drag effect in both modes we derive here will not appear in a %non-relativistic fluid.

\subsection{Relevance to physical systems}
To connect the dimensionless dispersion relation and numerical results to physical systems, we first specify the radiation field measured in the unperturbed fluid frame. For a source of luminosity $L_{\rm rad}$, the lab-frame radiation energy density at distance $r$ is
\begin{equation}
U_{\rm lab} = \frac{L_{\rm rad}}{4\pi r^2 c}.
\label{eq:Ulab_general}
\end{equation}
%For the plane-parallel geometry adopted in the derivation above, the corresponding comoving radiation energy density is
%\begin{equation}
%U'_{\rm phys} = \Gamma_0^2(1-\beta_0)^2 U_{\rm lab},
%\qquad
%\beta_0=\sqrt{1-\Gamma_0^{-2}}.
%\label{eq:Uprime_general}
%\end{equation}
The corresponding comoving radiation energy density is $U' = \Gamma_0^2(1-\beta_0)^2 U_{\rm lab}$ for a streaming radiation field propagating along the same axis as the bulk flow, and $U'=\Gamma_0^2(1+\beta_0^2/3)U_{\rm lab}$ for an isotropic radiation field.

The natural inverse-Compton interaction rate associated with $U'$ is
\begin{equation}
A_{\rm phys} = \frac{\sigma_{\rm T}U'}{m_ec}.
\label{eq:Aphys_general}
\end{equation}
Its inverse, $A_{\rm phys}^{-1}$, sets the basic Thomson radiation-interaction timescale. The actual radiative cooling time of the electrons additionally depends on their energy distribution and can be obtained directly from the time component of the radiation four-force, Eq.~\eqref{eq:energy_term_force}. Since $p'^0=E'_e/c$, the mean inverse-Compton energy-loss rate is
\begin{equation}
\left\langle\frac{dE'_e}{dt'}\right\rangle =
\frac{c f^0}{n_e} =
-\frac{4}{3}\sigma_{\rm T}cU'_{\rm phys} \left\langle\gamma^2\beta^2\right\rangle .
\label{eq:IC_loss_rate}
\end{equation}
Taking the mean electron kinetic energy,
$\left\langle E'_{\rm kin}\right\rangle
=
m_ec^2\left\langle\gamma-1\right\rangle$,
the characteristic radiative-loss time is
\begin{equation}
t'_{\rm loss}
\equiv
\frac{\left\langle E'_{\rm kin}\right\rangle}
{\left|\left\langle dE'_e/dt'\right\rangle\right|}
=
\frac{3}{4A_{\rm phys}}
\frac{\left\langle\gamma-1\right\rangle}
{\left\langle\gamma^2\beta^2\right\rangle}.
%\\
%&=
%\frac{3m_ec}{4\sigma_{\rm T}U'_{\rm phys}}
%\frac{\left\langle\gamma-1\right\rangle}
%{\left\langle\gamma^2\beta^2\right\rangle}
%\nonumber\\
\label{eq:trad_general}
\end{equation}
For an ultra relativistic electron population, $\gamma\gg1$ and $\beta\simeq1$, this reduces to
\begin{equation}
t'_{\rm loss}(\gamma)
\simeq
\frac{3m_ec}
{4\sigma_{\rm T}U'_{\rm phys}\gamma}.
\label{eq:trad_ultrarel}
\end{equation}

%In the worked examples below we usually specify the mode by the dimensionless ratio $k/\omega_0$. 
%In that form,
%\begin{equation}
%k_{\rm phys}
%=
%\left(\frac{k}{\omega_0}\right)
%\frac{\omega_{0,\rm phys}}{c},
%\qquad
%\lambda_{\rm phys}
%=
%\frac{2\pi c}
%{(k/\omega_0)\,\omega_{0,\rm phys}}.
%\label{eq:lambda_plateau_general}
%\end{equation}

For a perturbation proportional to $\exp(-i\omega t)$, 
the physical e-folding time for growth or damping is
\begin{equation}
\tau_{\rm rad}
=
\frac{1}{|\operatorname{Im}\omega_{\rm phys}|}
%=
%\frac{t_{\rm unit}}
%{|\operatorname{Im}\omega_{\rm code}|}.
\label{eq:taue_general}
\end{equation}
which depends on the ratio $f_0/w_0$ as derived above. 

Generally, neither $n_{\rm e}$ nor $w_0$ can be easily determined for physical systems. However, we can write this ratio entirely in terms of $U'$ (which can be reasonably estimated) and the set of dimensionless parameters defined in eqs.~(\ref{eq:chi}) and (\ref{eq:tildechi}) above:
\begin{equation}
    \omega_0=\frac{f_0}{w_0}=\frac{\sigma_{\rm T}U'_{\rm phys}}{m_{\rm e}}\tilde{\chi}
\end{equation}
which can be evaluated for a range of different assumptions about the proton content of the fluid and the electron spectrum implicit in $\tilde{\chi}$.

Next, we apply the results to three astrophysical systems: quasar jets, GRBs and blazars.

\subsubsection{Example: quasar 3C 279}
\label{sec:3C279}
First, we look at the BLR-dominated epoch of the flat-spectrum radio quasar 3C~279 analyzed by \citet{thekkoth_understanding_2023} for MJD~56642-56649, characterized by $\Gamma_0=14.05$, $\beta_0=0.99746$, and an external radiation energy density
\begin{equation}
    U_{\rm lab}=7.06\times10^{-3}\ {\rm erg\,cm^{-3}},
\end{equation}
taken from their broadband SED fit (Table 9, EC/BLR epoch). The size of the BLR is estimated to be $R=10^{16}$ cm and the estimate $r_{\rm diss}\sim \Gamma_0 R \sim 1.4\times 10^{17}$cm which is within the $R_BLR\sim 0.01$pc for bright disk. The BLR photon field is therefore almost isotropic, and the comoving energy density is boosted
\begin{equation}
    U'_0 = \Gamma_0^2\!\left(1+\tfrac{1}{3}\beta_0^2\right)U_{\rm lab} = 1.86\ {\rm erg\,cm^{-3}}.
\end{equation}
We will use this value to estimate the relevant timescales below.
If the jet is traveling outside the BLR, the radiation field would be diminished by a factor $\sim(R_{\rm BLR}/D)^2$, in case the boosted field will approach the point source transformation $\Gamma_0^2(1-\beta_0)^2 U_{\rm lab}$. %which does not apply here.

The non-thermal electron population is described by a broken power law 
\begin{equation}
N(\gamma)\,d\gamma \propto
\begin{cases}
\gamma^{-p}\,d\gamma, & \gamma_{\min}\le\gamma\le\gamma_{\rm b},\\[4pt]
\gamma_{\rm b}^{\,q-p}\,\gamma^{-q}\,d\gamma, & \gamma_{\rm b}<\gamma\le\gamma_{\max},
\end{cases}
\label{eq:broken_powerlaw}
\end{equation}
with $\gamma_{\rm min}=40$, $\gamma_{\rm b}=750$, $\gamma_{\rm max}=10^7$, $p=1.961$, $q=4.687$. 

Equation \eqref{eq:Aphys_general} gives 
\begin{equation}
A_{\rm phys} = \frac{\sigma_{\rm T}U'_0}{m_ec} = 4.52\times10^{-8}\ {\rm s^{-1}}.
\end{equation}
The moments entering the radiation force are
\begin{equation}
\left\langle\gamma^2\beta^2\right\rangle_0 = 5.10\times10^4,
\qquad
\left\langle\gamma-1\right\rangle_0 = 1.41\times10^2,
\end{equation}
which gives $\xi=0.999971$, and both modes will be damped.

In the cold relativistic jet case, $P_0/\rho_0=10^{-2}$, Eq.~\eqref{eq:xihigh} gives $\xi_{\rm high}=0.3781$, so the population lies well inside the high-$\xi$ unstable region.

The radiative-loss time using Eq.~\eqref{eq:trad_general},
\begin{align}
t'_{\rm loss}
&= \frac{3}{4A_{\rm phys}}\frac{\left\langle\gamma-1\right\rangle}{\left\langle\gamma^2\beta^2\right\rangle}
\nonumber\\
&= 4.57\times10^4\ {\rm s} \simeq 12.7\ {\rm hr}.
\label{eq:3c279_trad}
\end{align}

%From Eqs.~\eqref{eq:omega0_from_xi}-\eqref{eq:Imomega_phys_general}
%\begin{equation}
%\begin{aligned}
%\omega_{0,\rm phys} &= 7.14\times10^{-9}\ {\rm s^{-1}} ,\\
%\left(\frac{\mathrm{Im}\,\omega}{\omega_0}\right)_{k\to\infty} &= 1.831 ,\\
%\operatorname{Im}\omega_{\rm phys} &= 1.31\times10^{-8}\ {\rm s^{-1}}.
%\end{aligned}
%\end{equation}
The broadband SED fit from \citep{thekkoth_understanding_2023} uses a BLR  emission-region size of $R\sim10^{16}\ {\rm cm}$, and viewing angle $\theta=2^\circ$. 

The dynamical time of the jet inside the BLR is
\begin{equation}
t'_{\rm dyn} = \frac{R}{\Gamma_0c} = 2.3\times10^{4}\ {\rm s} = 6.6 \ {\rm hours}.
\end{equation}

The e-folding time, from Eq.~\eqref{eq:taue_general}, depends on the composition factor $\chi=(1+m_p n_p/m_e n_e)^{-1}$, ranging from $\chi=1$ for a pure pair plasma to $\chi=m_e/m_p$ for a cold proton-dominated jet, and the dimensionless enthalpy $\tilde{w_0}$. Table~\ref{tab:3c279_chi} lists the resulting times for a cold ($P_0/\rho_0=10^{-2}$, $\tilde{w}_0=1.04$; growing) and a hot ($P_0/\rho_0=10$, $\tilde{w}_0=41$; damping) state for both compositions.

\begin{table}[t]
\centering
\begin{tabular}{cccl}
\hline
$P_0/\rho_0$ & $\tilde{w}_0$ & $\chi$ & $\tau_e$ \\
\hline
$10^{-2}$ & $1.04$ & $1$        & $6.2$~min (grow) \\
$10^{-2}$ & $1.04$ & $m_e/m_p$  & $7.85$~days (grow) \\
$10$      & $41$   & $1$        & $7.51$~hr (damp) \\
\textcolor{gray}{$10$}      & \textcolor{gray}{$41$}   & \textcolor{gray}{$m_e/m_p$}  & \textcolor{gray}{$1.57$~yr (damp)} \\
\hline
\end{tabular}
\caption{Damping/growth times $\tau_{\rm rad}$} for the 3C 279 at two thermodynamic states $\tilde{w}_0$ and two compositions $\chi$.
\label{tab:3c279_chi}
\end{table}

For a pair-dominated jet in the cold state with a non-thermal tail, $\tau_{\rm rad}\simeq6.2$ min so
the three characteristic timescales therefore is
\begin{equation}
\tau_{\rm rad} \ll t'_{\rm loss} \ll t'_{\rm dyn}.
\end{equation}

The growth time is thus two orders of magnitude shorter than the region's light-crossing time, and two orders of magnitude shorter than the radiative loss time. Because this electron population lies in the unstable region, radiation-drag induced growth/damping is fast and dynamically important during this BLR-dominated epoch.
For a proton-dominated jet in the cold state, $\tau_{\rm rad}\simeq 7.85$ days which is larger than the $t'_{\rm dyn}$ now, so the effect of the perturbation is small before it goes out of the region. In the cold state, both compositions give damping modes.

With the same 3C 279 electron population and radiation field above, we can find the critical pressure-to-density ratio that determines the growing mode and damping mode.
Solving $\xi_{\rm high}(P_0/\rho_0)=\xi$ in closed form from the quadratic $a_s^2+a_s-\xi/3=0$ and $a_s^2=(4/3)(P_0/\rho_0)/[1+4(P_0/\rho_0)]$ for $\Gamma_{\rm ad}=4/3$ gives
\begin{equation}
\begin{aligned}
a_{s,\rm crit} &= \frac{-1+\sqrt{1+4\xi/3}}{2} = 0.2638, \\
\left(\frac{P_0}{\rho_0}\right)_{\rm crit} &= \frac{a_{s,\rm crit}^2}{4/3-4a_{s,\rm crit}^2} = 0.0659
\end{aligned}
\end{equation}
below which the mode grows and above which it damps.
The cold case ($P_0/\rho_0=10^{-2}$) lies well below $(P_0/\rho_0)_{\rm crit}$ and grows, as found above, while the hot cases $P_0/\rho_0 =10$ damp.

Table~\ref{tab:examples_summary} also gives the damping times for two hotter compositions, $P_0/\rho_0=1$ and $10$, representative of a strong internal shock or a pair-dominated outflow. In summary, for cold pair dominated jets , radiation drag can be important and the dispersion relation shows that under such circumstances, sound waves can be modified by damping/growth within the BLR. For cold proton-dominated jets or hotter pair dominated jets, the effect can be marginally growing or damping, and for hotter proton-dominated jets the effect is negligible.

\begin{table*}
  \centering
  \caption{Summary of the radiation-drag growth and damping timescales derived for the worked examples of Section~\ref{sec:discussion} for $\chi =1$ (pair-dominated plasma). "iso." and "stream." denote the isotropic and streaming boost transforms. Timescales are e-folding ($\tau_e$, growing) or damping ($\tau_{\rm damp}$) times. Gray text denotes scenarios where radiation drag is insignificant.}
  \label{tab:examples_summary}
  \begin{tabular}{llcccc}
    \toprule
    Source & Radiation field (boost) & $\Gamma_0$ & $U'$ [erg\,cm$^{-3}$] & $\tau_e$ or $\tau_{\rm damp}$   \\
    \hline
    3C~279            & BLR (iso.), $P_0/\rho_0=10^{-2}$      & 14.05 & 1.86  & 6.2 min              \\
                       & BLR (iso.), $P_0/\rho_0=1$             & 14.05 & 1.86  & 1.04 hr (damp)                         \\
                       & BLR (iso.), $P_0/\rho_0=10$            & 14.05 & 1.86  & 7.51 hr (damp)                          \\
    \hline
    GRB~120709A        & \textcolor{gray}{progenitor WR wind (stream.)}                          & \textcolor{gray}{150}   & \textcolor{gray}{$1.18\times10^{-3}$}  & \textcolor{gray}{$3.48\times10^{-3}$~yr}             \\
                       & O-star companion, $d{=}3$~AU (iso.)                 & 150   & $1.98\times10^{3}$ & $6.55\times10^{-2}$~s                  \\
    \hline
    PKS~J1421-0643     & CMB, $z=3.689$ (iso.)                       & 4     & $4.24\times10^{-9}$  & $1.84\times10^{4}$~yr  \\
                       & \textcolor{gray}{host elliptical (iso.)}                      & \textcolor{gray}{4}     & \textcolor{gray}{$6.72\times10^{-11}$} & \textcolor{gray}{$1.16\times10^{6}$~yr} \\
                       & CMB, $P_0/\rho_0=1$ (iso.)                  & 4     & $4.24\times10^{-9}$  & $1.9\times10^{5}$~yr (damp)  \\
    \hline
  \end{tabular}
\end{table*}

\subsubsection{Example: GRB 120709A}
We apply the instability analysis to GRB~120709A studied by \citet{bukhari_spectral_2022} via \textit{Fermi} GBM-LAT spectroscopy, using their representative jet composition and electron population from the first emission episode ($\Gamma_0=150$, $\gamma_{\rm min}=10^2$, $\gamma_{\rm max}=10^4$, $p=2.5$), for which Eq.~\eqref{eq:xihigh} gives $\xi=0.99999\gg\xi_{\rm high}=0.3781$ at $P_0/\rho_0=10^{-2}$.

Setting $\xi_{\rm high} = \xi$ gives a critical ratio $(P_0/\rho_0)_{\rm crit} \approx 0.066$, above which the mode is damped. A hot plasma jet with $P_0\gg \rho_0$ would fall into the purely damped regime. The cold value adopted here ($P_0/\rho_0 = 10^{-2}$) is representative of this example in the growing regime (this again could either be a cold thermal core with a nonthermal tail, or proton-dominated). 

The instability grows on the time scale $\tau_\omega=1/\mathrm{Im}(\omega)_{\rm phys}$, which we compare with the dynamical time $t_{\rm dyn}=r/(\Gamma_0 c)$ at the distance $r$ of the radiation source from the jet.
For reference, at the MeV emission radius $R_{\rm MeV}=3\times10^{14}\ {\rm cm}$ the burst dynamical time is $t'_{\rm dyn}=R_{\rm MeV}/(\Gamma_0 c)=66.7\ {\rm s}$, and the comoving density needed to e-fold within it is
\begin{align}
    U'_{\rm crit} &= \frac{m_ec}{\sigma_T\,t'_{\rm dyn}}\cdot\frac{w_0}{\left(1+\tfrac{2}{3}\langle\gamma^2\beta^2\rangle_0\right)\left(\mathrm{Im}\,\omega/\omega_0\right)_{k\to\infty}} \\
    &= 1.94\ {\rm erg\,cm^{-3}}.
\end{align}

Converting $U'_{\rm crit}$ into a required host-frame luminosity at $r=R_{\rm MeV}$ depends strongly on the assumed source geometry, which enters through the Doppler boost factor $\mathcal{B}$ relating the lab- and comoving frame energy densities, $U'=\mathcal{B}\,U_{\rm lab}$. We consider two possibilities: a wind or envelope left over from the precursor along the jet axis $\mathcal{B}_{\rm stream}=\Gamma_0^2(1-\beta_0)^2$, and a stellar companion in orbit with the precursor $\mathcal{B}_{\rm iso}=\Gamma_0^2(1+\beta_0^2/3)$.
So the time ratio depends on 
\begin{equation}
\frac{\tau_\omega}{t_{\rm dyn}}
= \frac{4\pi m_e c^3\,\tilde{w}_0\,\Gamma_0}
{\sigma_T\,\mathcal{B}\,\bigl[1+\tfrac23\langle\gamma^2\beta^2\rangle_0\bigr]\,(\mathrm{Im}\,\omega/\omega_0)_{k\to\infty}}
  \,\frac{r}{L}
\label{eq:grb_ratio}
\end{equation}

For the axial wind or envelope behind the emission with $\mathcal{B}_{\rm stream}=\Gamma_0^2(1-\beta_0)^2\simeq1.1\times10^{-5}$,
\begin{equation}
\left.\frac{\tau_\omega}{t_{\rm dyn}}\right|_{\rm axial}
\approx 9\times10^{6}\left(\frac{r}{3\,{\rm AU}}\right)\!\left(\frac{10^{38}\,{\rm erg\,s^{-1}}}{L}\right).
\end{equation}
A typical Wolf-Rayet star \citep{crowther_physical_2007} with luminosity $\sim10^{39}\ {\rm erg\,s^{-1}}$ at $r\sim R_{\rm MeV}$ gives $\tau_\omega/t_{\rm dyn}\sim6\times10^{6}$, so a source the along the axis with stream boost cannot drive any meaningful perturbation growth or damping.

Most massive stars form in binaries, and their companions are frequently massive themselves, so it is plausible for a core-collapse GRB progenitor to be orbited by an O-star with luminosity $L_\star\sim$~few$\times10^{37}$-$10^{38}\ {\rm erg\,s^{-1}}$ \citep{martins_new_2005}  at an orbital separation of a few AU \citep{sana_binary_2012}. Since the companion's direction relative to the jet axis varies with orbital phase rather than lying fixed along it, we treat this as an ambient, direction-independent field and apply the isotropic boost $\mathcal{B}_{\rm iso}=\Gamma_0^2(1+\beta_0^2/3)\simeq3.0\times10^{4}$,
\begin{equation}
\left.\frac{\tau_\omega}{t_{\rm dyn}}\right|_{\rm companion}
\approx 3.3\times10^{-3}\left(\frac{r}{3\,{\rm AU}}\right)\!\left(\frac{10^{38}\,{\rm erg\,s^{-1}}}{L}\right).
\end{equation}
At the fiducial $L_\star=5\times10^{37}\ {\rm erg\,s^{-1}}$ and $r=3$~AU the mode e-folds $\sim150$ times within the dynamical time, and radiation drag remains effective for binary separations up to
\begin{equation}
r\lesssim 9\times10^{2}\left(\frac{L_\star}{10^{38}\,{\rm erg\,s^{-1}}}\right){\rm AU}.
\end{equation}

\subsubsection{Example: PKS J1421-0643}
We apply the same analysis to the blazar PKS~J1421-0643 \citep{worrall_inverse-compton_2020}, using their Jet~D parameters
($z=3.689$, $\Gamma_0=4$, $\theta=14.5^\circ$, $\delta=4.0$, $B=2.95\ {\rm nT}$, $\gamma_{\rm min}=10$, $\gamma_{\rm max}=5000$, $p=2.3$, $R_{\rm deproj}=130\ {\rm kpc}$), for which Eq.~\eqref{eq:xihigh} gives $\xi=0.99989\gg\xi_{\rm high}=0.3781$ at $P_0/\rho_0=10^{-2}$ ($w_0 = 1.04$). The lab-frame CMB energy density at this redshift is $U_{\rm CMB}(z)=2.02\times10^{-10}\ {\rm erg\,cm^{-3}}$, boosted isotropically to $U'_{\rm CMB}=4.24\times10^{-9}\ {\rm erg\,cm^{-3}}$. 

%From Eqs.~\eqref{eq:omega0_from_xi}-\eqref{eq:Imomega_phys_general}
%%\begin{equation}
%\begin{aligned}
%\omega_{0,\rm phys} &= 9.40\times10^{-13}\ {\rm s^{-1}}\\
%\left(\frac{\mathrm{Im}\,\omega}{\omega_0}\right)_{k\to\infty} &= 1.830\\
%\mathrm{Im}(\omega)_{\rm phys} &= 1.72\times10^{-12}\ {\rm s^{-1}}
%\end{aligned}
%\end{equation}
The e-folding time, from Eq.~\eqref{eq:taue_general}, is
\begin{equation}
\tau_e = 5.81\times10^{11}\ {\rm s} = 1.84\times10^{4}\ {\rm yr}.
\end{equation}
The comoving dynamical time corresponding to the measured deprojected jet length is 
\begin{equation}
    t'_{\rm dyn} = \frac{R_{\rm deproj}}{\Gamma_0 c} = 3.35\times10^{12}\ {\rm s} = 1.06\times10^{5}\ {\rm yr}
\end{equation}
where the $\Gamma_0$ in the denominator accounts for both time dilation and Lorentz contraction in converting from the lab frame to the comoving jet frame, 
while the corresponding lab-frame light-travel time $\frac{R_{\rm deproj}}{\beta_0 c} = 4.38\times10^{5}\ {\rm yr}$.
%\begin{equation}
%\frac{R_{\rm deproj}}{\beta_0 c} = 1.38\times10^{13}\ {\rm s} = 4.38\times10^{5}\ {\rm yr}
%\end{equation}
And the minimum propagation length required for one e-fold gives $R_{\rm crit}= 22.6\ {\rm kpc}$
%\begin{equation}
%R_{\rm crit} = \Gamma_0 c\,\tau_e = 6.97\times10^{22}\ {\rm cm} = 22.6\ {\rm kpc}
%\end{equation}
which is smaller than $R_{\rm deproj}$, so this setup is plausible for the resulting dispersion to be physically relevant over the jet's length.

Next, if we consider the same jet passing through a separate, typical elliptical galaxy radiation field, $R_{\rm gal} = 10\ {\rm kpc}$ with $L_{\rm gal} = 10^{11}\ L_\odot = 3.83\times10^{44}\ {\rm erg\,s^{-1}}$.
The energy density is $U_{\rm gal} = 3.20\times10^{-12}\ {\rm erg\,cm^{-3}}$.
Suppose the diffuse field has no preferred direction, so we apply the isotropic field boost $U'_{\rm gal} = 6.72\times10^{-11}\ {\rm erg\,cm^{-3}}$.
%\begin{equation}
%U'_{\rm gal} = \Gamma_0^2\!\left(1+\frac{\beta_0^2}{3}\right)U_{\rm gal} = 6.72\times10^{-11}\ {\rm erg\,cm^{-3}}
%\end{equation}

%From Eqs.~\eqref{eq:omega0_from_xi}--\eqref{eq:Imomega_phys_general}
%\begin{equation}
%\begin{aligned}
%\omega_{0,\rm phys} &= 1.49\times10^{-14}\ {\rm s^{-1}}\\
%\left(\frac{\mathrm{Im}\,\omega}{\omega_0}\right)_{k\to\infty} &= 1.830\\
%\mathrm{Im}(\omega)_{\rm phys} &= 2.73\times10^{-14}\ {\rm s^{-1}}
%\end{aligned}
%\end{equation}
The e-folding time, from Eq.~\eqref{eq:taue_general}, is
\begin{equation}
\tau_e = 3.66\times10^{13}\ {\rm s} = 1.16\times10^{6}\ {\rm yr}.
\end{equation}

Using the same comoving dynamical time as the CMB example, $t'_{\rm dyn}=1.06\times10^{5}\ {\rm yr}$,
the minimum propagation length for one e-fold is $R_{\rm crit} = 1423\ {\rm kpc}$
%\begin{equation}
%R_{\rm crit} = \Gamma_0 c\,\tau_e = 4.39\times10^{24}\ {\rm cm} = 1423\ {\rm kpc}
%\end{equation}
about $11\times$ longer than the jet's actual observed length and two orders of magnitude larger than the galaxy (and thus the scale for the radiation field.) This galaxy will contribute much less to the radiation drag term than the CMB radiation and we can neglect its effects on the dispersion relation.

%As a check on the boost geometry, applying instead the streaming boost transform $U'=\Gamma_0^2(1-\beta_0)^2\,U_{\rm gal}$ gives
%\begin{equation}
%\begin{aligned}
%U'_{\rm gal,\ stream} &= 5.17\times10^{-14}\ {\rm erg\,cm^{-3}} ,\\
%\tau_{e,\rm stream} &= 4.77\times10^{16}\ {\rm s} = 1.51\times10^{9}\ {\rm yr} ,\\
%R_{\rm crit,\ stream} &= 5.72\times10^{27}\ {\rm cm} = 1.85\times10^{6}\ {\rm kpc} \\ & = 1852\ {\rm Mpc} .
%\end{aligned}
%\end{equation}

To test the damped region, for a hotter composition,  $P_0/\rho_0=1$ applied to the same electron population and the CMB field, the mode is damped with $\tau_{\rm damp}\approx1.9\times10^{5}\ {\rm yr}$, comparable to the growth found in the unstable case above. Therefore, radiation drag affects the flow regardless of whether the jet is in the growing mode or the damped mode.

Finally, we note that the CMB energy density scales with redshift $U_{\rm CMB}(z)= U_{\rm CMB,0} (1+z)^4$. %, so the CMB contribution to radiation drag rises by four powers of $(1+z)$. 
High-redshift blazars and any other ultra-relativistic sources at high $z$ are more strongly affected by CMB radiation drag than low-redshift sources, so the CMB field is strong enough to modify wave propagation.

\subsection{\label{sec:limitations}Caveats}
Our analysis is done in localized conditions in order to study small perturbations. There are several limitations of the work in this manuscript which we discuss below:

\subsubsection{Locality and Causality.}The radiation force and hydrodynamic equations are derived in the initial rest frame of the fluid. In this frame, the initial unperturbed flow is at rest, and the radiation force appears as an external source term that accelerates/decelerates the flow. The dispersion relations are also evaluated using the initial background pressure and density. The operating assumption is that we can define such a rest frame where the fluid can be treated as causally connected over several wavelengths and that fluid properties do not change materially over an oscillation. 

%Thus, our analysis is only valid for a local short-time interpretation, assuming the acceleration is small and does not affect the mean fluid velocity, pressure, and density. Our derivations do not work if the flow has accelerated substantially, when we cannot approximate the conditions of the fluid with the initial conditions, as shown in the later times of the simulation plots. 
When the background velocity $\beta_0$ is no longer small, and relativistic effects become important, simultaneity effects also become important and longer-wavelength perturbations go out of causal contact, invalidating our assumption that we can transform to the inertial rest frame of the fluid and treat the resulting equations as non-relativistic.

\begin{figure}
    \begin{center}  
    \includegraphics[width=\linewidth]{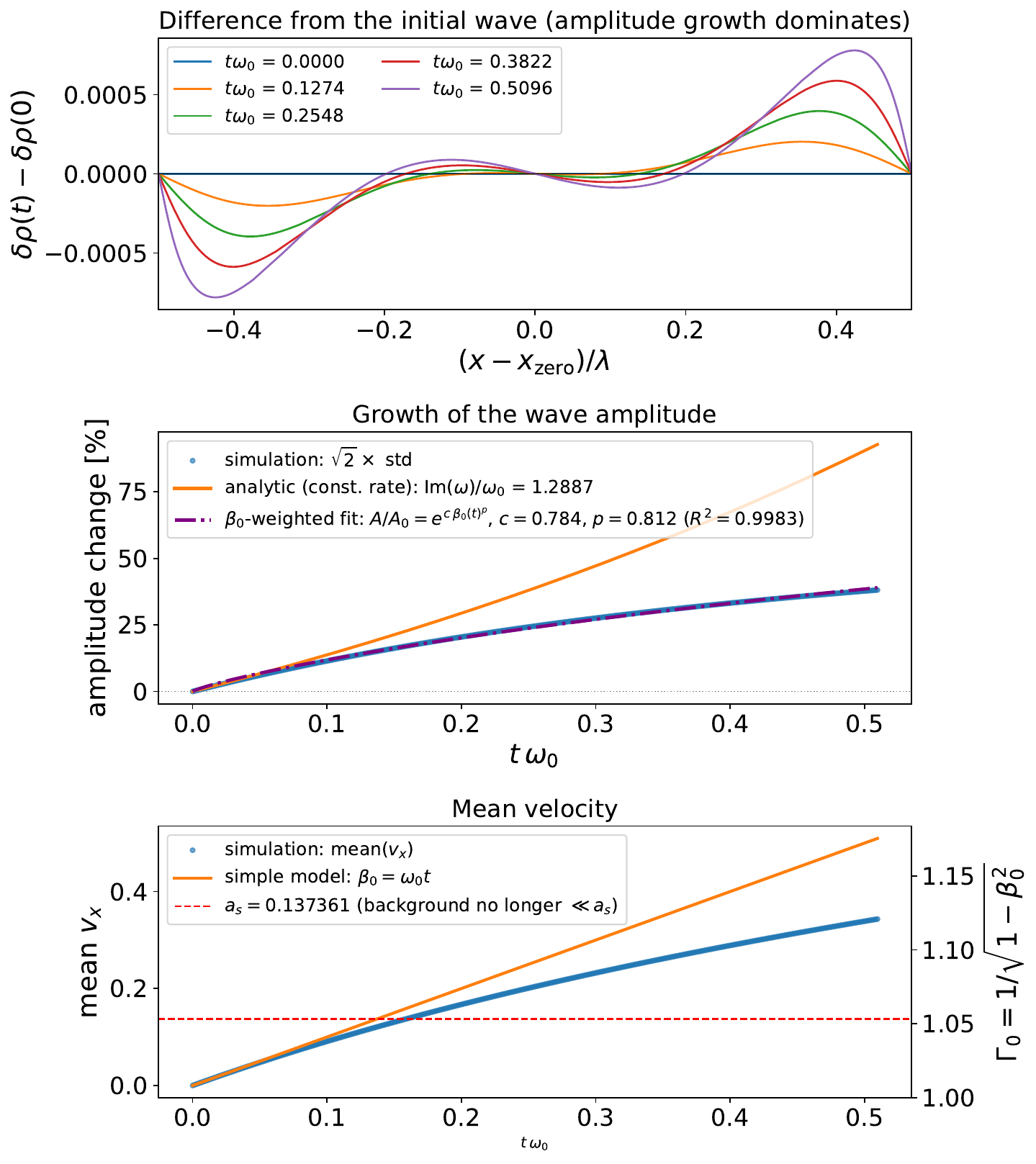}
    \end{center}
    \caption{Test~1, $\omega_+$ (growing) mode, run to late times, beyond the regime in which this local, short-time approximation is hold. \textit{Top:} the perturbed density profile at selected times, aligned on their first zero crossing and subtracted the initial $t\omega = 0$ profile. The growing deviates from the simple sine function as in plot (d) of Fig. \ref{fig:Test1_linear_validation}. \textit{Middle:} percent change in wave amplitude versus time. The blue points are the amplitude measured directly from the simulation ($\sqrt{2}\times$ the standard deviation off the perturbed density); the solid orange line is the analytic prediction assuming a fixed, stationary background, $\mathrm{Im}(\omega)/\omega_0=1.2887$; the dash dotted purple line is a two parameter fit of the form $A(t)/A_0=\exp[c\,\beta_0(t)^p]$, using the measured background velocity $\beta_0(t)$ itself (bottom panel) which recovers the simulated growth to $R^2=0.998$ with best-fit $c=0.784$ and $p=0.812$. \textit{Bottom:} mean flow velocity $\beta_0=\langle v_x\rangle$ versus time. Points are the simulated mean velocity; the solid orange line is the naive linear model $\beta_0=\omega_0 t$; the dashed red line marks the fluid sound speed $a_s=0.137$, beyond which the background can no longer be treated as approximately at rest; the right-hand axis shows the corresponding bulk Lorentz factor $\Gamma_0=(1-\beta_0^2)^{-1/2}$. }
    \label{fig:longrun}
\end{figure}

Fig.~\ref{fig:longrun} shows this for Test 1's growing mode, run past the timescale where the assumption holds. The bottom panel tracks the mean flow velocity $\beta_0(t)$ and plots the naive constant force model $\beta_0=\omega_0t$. They agree at early times but diverge once $\beta_0$ is large (at about the sound speed $a_s$), and the radiation force itself weakens as $(1-\beta_0)^2/\Gamma^3$. The middle panel shows the corresponding effect on the wave amplitude. The measured growth falls well below the analytic prediction $\mathrm{Im}(\omega)/\omega_0=1.2887$. The top panel confirms that the perturbation itself still grows, but the wave is no longer the single, pure sinusoidal mode that we seeded at the beginning for the linear dispersion relation and other harmonics start to appear (see \S\ref{sec:linearity}).

This limitation also affects the plane-parallel treatment of the radiation field that we discussed at the beginning of Section \ref{sec:linear_analysis}. For a field intrinsically isotropic but seen by a fluid with $\Gamma\gg1$ (regime~b), the comoving energy density $U'$ is evaluated at the initial Lorentz factor $\Gamma_0$. In a fully nonlinear simulation where $\Gamma$ is allowed to evolve, this update would follow from the $\Gamma$ dependence of the drag four-force. Our present runs hold the background fixed at $\Gamma_0$ in  both the analytic and numerical treatments. We leave the time-dependent treatment as one of the future works in Section \ref{sec:conclusion}.

\subsubsection{Linearity.}\label{sec:linearity}  Our calculation is restricted to linear order in all the perturbed terms in Eq.~\ref{eq:pertubation}, ignoring all higher-order terms. The resulting analytic dispersion relation describes only the small-amplitude modes that we seeded, and does not contain any nonlinear instability or interference between Fourier modes. This is particularly important as in the numerical simulation comparison, the analytic growth or damping rates should only be compared with early time intervals when the perturbation remains close to a single linear mode. At later times, the nonlinear terms and interferences among different modes may change the results, which will be interesting to study.

\subsubsection{One-dimensional perturbations.} \label{sec:onedimcaveat} Our dispersion relation is derived for plane-wave perturbations along a single direction, and does not include transverse or oblique modes, curvatures, or any three-dimensional structure of the flow. Real jets are not infinite or uniform in one direction alone, and real-world boundary effects will likely change the final dispersion. Our results should stay valid as long as the assumptions of a local fluid inside the jet and the short-time analysis are met, that is, for short wavelength modes, which we have identified above as the most relevant already. Allowing oblique modes is still tractable in principle but introduces the full multi-dimensional structure. 

\subsubsection{External radiation field geometry.} The magnitude of the radiation field $U'_0$ depends on the geometry of the external field relative to the jet's motion. We considered two idealized cases, a streaming, point-source field behind the fluid, for which $U'_0=\Gamma_0^2(1-\beta_0)^2U_{\rm lab}$, and an isotropic field, for which $U'_0=\Gamma_0^2(1+\beta_0^2/3)U_{\rm lab}$. Real astrophysical photon fields are not perfect, with anisotropic angular distributions. Thus, our examples with the specific numbers are only estimates from idealized, locally homogeneous backgrounds.

\subsubsection{No radiative feedback.} Our local model also does not include the radiation field change from the fluid itself. We treat $U'$ as an external, uniform radiation field which is also affected by the kinematically perturbed fluid $\delta U' = 2 U'_0 \delta \beta'$. This model that only the external radiation acts on the fluid, but the fluid's radiation is not taken into account, is a good approximation when the fluid is optically thin. In an optically thick environment where the electrons in the fluid are a non-negligible source of the radiation field and where not all of the fluid may be subject to the external radiation field, a fully coupled radiation-hydrodynamic system would be required. Thus, our results will be limited to cases where the wavelength is small compared to a photon mean free path.

\subsubsection{Pure hydrodynamics.} We include the radiation drag force term as an external force term in the pure hydrodynamic equations. There is no magnetic field. The fluid pressure and the stability boundaries are derived only for the gas pressure. In the case of magnetic-dominated jets, a relativistic MHD extension of this formalism would be needed to modify the stability analysis accordingly.

\subsubsection
{No radiative cooling of electrons.}\label{sec:nocoolingcaveat} In the dispersion relation, the $\langle\gamma^2\beta^2\rangle_0$ term from the electron population is fixed throughout the analysis, but in reality the high-energy electrons will cool due to inverse Compton cooling, and the $\gamma^2$ term will decrease over time, self-limiting the contribution of this term to radiation drag in conditions where the radiative cooling time of electrons is short compared to the acceleration time. This is a generic feature of radiation drag on relativistic particles, not specific just to the present formalism.

While of general concern when discussing radiation drag effects, there may be astrophysically relevant cases where the radiative cooling time is longer than the acceleration time. For example, in the 3C 279 case study discussed in \S\ref{sec:3C279}, the radiative cooling time is much longer than the instability growth time, so the electron population does not cool significantly over one growth time.

Radiative cooling also does not affect the bulk drag contribution of $\omega_0$. 

Also, since radiative losses affect the efficiency of radiation drag of the mean flow by the same factor, this caveat does  not change our conclusion that whenever radiation drag itself is important for the background flow, its effects discussed in this paper on fluid perturbations are also important.

\subsubsection{No dissipation.}
Finally, our treatment does not include dissipative terms. Viscous damping will act in concert with radiation damping as discussed here, and in competition with radiative instability where present. Since viscous effects will, on average, increase particle energies, they will also counteract radiative cooling, however, heating will not preferentially affect the highest energy electrons/pairs, and will thus not in itself increase $\xi$. Since viscous effects are of second order, they will also not substantially modify the first order dispersion relation derived above.

\subsubsection{Collisionless plasmas and kinetic effects.}
Many systems of interest (AGN jets, GRB outflows) involve collisionless plasma. The rest-frame isotropy has no collisional or magnetic process to maintain it, holding only as an injection condition over short times, and our single-fluid coupling breaks down because the radiation force acts only on the leptons, so a lepton–ion drift can develop and drive kinetic instabilities. Capturing these effects requires a kinetic (Vlasov–Maxwell) treatment, which we leave to future work.

\section{\label{sec:conclusion}Conclusion and Future Work}
We have derived the linear dispersion relation for acoustic-type perturbations in a relativistic fluid subject to Compton radiation drag. Every mode is either damped or unstable, with rates of order the local bulk-acceleration rate $\omega_0$. Instability occurs in two separate wavenumber ranges, with a stable band between them determined by how the electron parameter $\xi$ compares to $\xi_{\rm low}=\tfrac{3}{2}a_s^2$ and $\xi_{\rm high}=3(a_s^2+a_s)$. Short-wavelength modes are essentially ordinary sound waves $v_p\to a_sc$, independent of $\xi$. At long wavelength the mode is radiation-driven, with both growth rate and phase speed set jointly by $\xi$ and $a_s$. We validated these rates against special-relativistic Athena++ simulations with agreement to a few percent across both branches.

We applied the relation to astrophysical flows such as a GRB and a blazar jet subject to CMB or external galaxy radiation fields, and find the predicted growth/damping timescale spans from seconds to a few million years. The same jet and field can lie on either side of the growth/damping boundary depending on the jet's thermodynamic state. These results show that Compton radiation drag instabilities and damping depend on the radiation environment and the jet composition. Because the growth and damping rates are of order the bulk radiative deceleration rate $\omega_0$, these perturbative effects become dynamically important only when radiation drag itself is dynamically important for the background flow.

The idealizations of this work discussed in Section~\ref{sec:limitations} point to several natural directions for future work: (i) following the instability into the nonlinear regime, where background acceleration and higher-order effects will eventually become important; (ii) extending the present relativistic hydrodynamic treatment to relativistic MHD with radiative transfer; (iii) evolving the electron distribution and electron-ion coupling self-consistently in kinetic or particle-in-cell calculations \citep{vanthieghem_role_2022}; (iv) considering realistic anisotropic radiation fields and fully three-dimensional flow geometries; (iv) studying real observables such as variability and polarization from the radiation drag instabilities.

\section{Acknowledgment}
We are grateful for comments on the manuscript from Ellen Zweibel and Karol Fulat. We further thank Mitch Begelman and Anatoly Spitkovski for helpful discussions and we thank Roark Habegger for help with the Athena++ setup and Puxin Lin for early discussions on the analytic formulation.
VZ acknowledges support from the National Science Foundation under NSF grant PHY-2409316, and the Department of Energy under the grant DE-SC0026099.

\appendix

\bibliography{Radiation_clean}
\end{document}